\documentclass[a4paper,fontsize=11pt,DIV=12,twocolumn=false]{scrartcl}

\usepackage[utf8]{inputenc}  
\usepackage[T1]{fontenc}  
\usepackage{lmodern}  
\usepackage{array}  
\usepackage{graphicx}  
\usepackage{float}  
\usepackage{amsmath}  
\usepackage{amssymb}  
\usepackage{amsfonts}  
\usepackage{textcomp}  
\usepackage{threeparttable}
\usepackage{enumitem}  
\usepackage{algpseudocode}  
\usepackage{algorithm}  
\usepackage{longtable}
\usepackage{pdflscape}
\usepackage{microtype}
\usepackage[labelfont=bf,font=small,format=plain,indention=0cm]{caption}  
\usepackage{authblk}  

\usepackage{booktabs}  
\usepackage{multirow}  
\usepackage[T1]{fontenc}
\usepackage{geometry}
\usepackage{setspace}

\usepackage{achemso}

\usepackage{graphicx}
\usepackage{float}
\usepackage{xcolor}
\usepackage{listings}
\usepackage{algorithm}
\usepackage{algpseudocode}
\newfloat{scheme}{htbp}{los}
\floatname{scheme}{Scheme}
\newfloat{graph}{htbp}{loh}
\floatname{graph}{Graph}

\usepackage{chemformula}
\usepackage[version = 4]{mhchem}
\usepackage{amsmath}
\usepackage{amsfonts}
\usepackage{amssymb}
\usepackage{bm}

\usepackage{xurl}
\usepackage{hyperref}

\definecolor{promptbg}{HTML}{F7F9FB}
\definecolor{promptframe}{HTML}{C8D2E0}
\lstdefinestyle{mchprompt}{
  basicstyle=\ttfamily\footnotesize,
  backgroundcolor=\color{promptbg},
  frame=single,
  rulecolor=\color{promptframe},
  framerule=0.4pt,
  framesep=4pt,
  breaklines=true,
  breakatwhitespace=false,
  columns=fullflexible,
  keepspaces=true,
  showstringspaces=false,
  xleftmargin=0.5em,
  xrightmargin=0.5em
}

\providecommand{\argmax}{\operatorname*{arg\,max}}

\usepackage{authblk}
\author[1]{César Ojeda\thanks{Corresponding author: \href{mailto:ojedamarin@uni-potsdam.de}{ojedamarin@uni-potsdam.de}}}
\author[2]{Darius A. Faroughy}
\author[3]{Maryam Karimi}
\author[4]{Payam Zarrintaj}
\author[3]{Mir Mehdi Seyedebrahimi}
\author[3]{Martín Carballo-Pacheco}
\affil[1]{\small Institute of Mathematics, Faculty of Science, University of Potsdam, Campus Golm, Building 9, Karl-Liebknecht-Str. 24–25, 14476 Potsdam-Golm, Germany}
\affil[2]{\small NHETC, Department of Physics and Astronomy, Rutgers University, Piscataway, NJ 08854, USA}
\affil[3]{\small Potsdam Transfer, University of Potsdam, Karl-Liebknecht-Str. 24-25, 14476 Potsdam.}
\affil[4]{\small Independent Researcher, E3 LLC, Louisville, KY, USA}

\title{My Chemical Harness: Evolutionary Molecular Design over Synthetic Pathways with Large Language Model Agents}
\date{} 

\begin{document}

\maketitle

\begin{abstract}
Designing molecules with target properties is most useful when candidate structures are accompanied by feasible synthetic routes. We introduce {\scshape My Chemical Harness}, a route-native evolutionary framework for goal-directed molecular design in which the search population consists of executable synthetic pathways rather than isolated molecular graphs. Each route is built from purchasable building blocks and reaction templates, executed by deterministic chemistry tools, and scored through task-specific molecular oracles. Large language models (LLMs) are used only as strategy controllers that select high-level preferences over route length, move type, reaction families, motifs, and exploration pressure, while local code performs route construction, validation, deduplication, scoring, selection, and memory updates. This separation lets the LLM guide exploration without allowing it to introduce hallucinated products or unsupported reaction steps. On a soluble epoxide hydrolase proxy task, our LLM agent improves over single pass LLM and deterministic controllers, reaching state-of-the-art performance across the sEH score, synthetic accessibility score, and AiZynthFinder success rate metrics. These results suggest that constrained LLM agents can play a significant role in molecular discovery without requiring training, fine-tuning, or dedicated generative models.
\end{abstract}

\section{Introduction}

A central challenge in drug and materials discovery is the identification and design of molecules with targeted chemical properties.
In light of recent advances in machine learning and generative modeling, whose applications have shown dramatic success in text
\cite{kojima2022large}, code \cite{austin2021program,chen2021evaluating}, image \cite{song2020score}, and video generation, increasing efforts have sought to tackle the
scientific discovery of new drugs and materials. To accomplish this task, and following standard practice, transformer based models as well as diffusion or 
transportbased methods have been directly applied to chemical data, yielding models that are able to sample new molecules in a \textit{de novo} 
manner. Despite promising successes, it has become clear that the direct application of these methodologies often yields candidates with a significant 
practical drawback: they may be synthetically inaccessible, meaning that no feasible synthesis route is known or readily available. Early 
attempts trained generative models directly on molecular representations such as SMILES strings \cite{qiu2024demand,gao2024generative,edwards2022translation} or 
3D molecular geometries \cite{hoogeboom2022equivariant,dunn2026flowmol3,song2023equivariant,vignac2022digress}, which contain no explicit 
information about the synthesizable routes required to obtain the  molecule. Moreover, because these models learn primarily from observed molecular distributions,
their ability to systematically explore beyond the support of the training data remains limited. Recent work has sought to incorporate synthetic pathways 
into molecular generation \cite{swanson2024generative,lee2026exploring, gao2024generative}, most commonly by projecting generated molecules onto synthesizable chemical space. In 
these approaches, synthesizability  is typically enforced a posteriori, as an auxiliary correction step applied on top of otherwise structurecentric 
generative methodologies.

It is worth observing that, once one focuses on available synthesizable routes, that is, on the set of reactions that can be reliably 
executed in the laboratory, one is confronted with a combinatorial search problem over reactions and purchasable building blocks. 
From this perspective, the chemist has access to a repository of meta programs, namely synthesizable route templates, which define 
reusable units of \textit{chemical computation} available for constructing candidate molecules. It is from this perspective that recent 
advances in large language models (LLMs) and agentic systems may offer substantial improvements for drug and material design. LLMs provide 
a powerful set of capabilities: their training on large scientific corpora enables them to generalize across chemical concepts, 
propose hypotheses for new routes, and operate at scale in repetitive and time consuming design loops. However, realizing this 
potential requires an inference time computational framework that allows LLMs to search chemical space under explicit constraints 
on budget, synthesizability, and target properties. Our task is therefore to design methodologies that equip LLMs with constraints, 
tasks, tools, and evaluation capabilities, collectively referred to as a harness, so that the resulting LLM based system can reliably
attain the desired design goals. In the present work, we propose a chemical harness that enables LLM agents to search synthesizable 
chemical space by combining synthetic route templates with purchasable building blocks. Inspired by recent successes in program search 
for coding and mathematics under the \textit{AlphaEvolve} paradigm\cite{romera2024mathematical,novikov2025alphaevolve}, we instantiate a 
genetic algorithm in which tool augmented LLM agents form the basis of the mutation operators. These operators evolve synthetic routes toward 
candidate molecules with desirable properties. We expect our contribution to facilitate the rational design of synthesizable drugs and molecularly imprinted polymers with high specificity for selected molecular targets.

Early attempts to use LLMs for materials design positioned the model primarily as a natural language interface or autonomous design assistant. 
ChatMOF \cite{kang2024chatmof} used LLMs to extract key information from user provided text and connect natural language queries to MOF prediction, 
retrieval, and generation workflows. LLMatDesign \cite{jia2024llmatdesign} extended this direction by using LLM agents to translate human
instructions, modify candidate materials, evaluate the outcomes using external tools, and refine subsequent decisions through self reflection. 
To improve such systems, emphasis has been placed on prompt engineering \cite{luo2025leveraging}, on agent roles that mimic a research group 
\cite{zheng2023chatgpt}, with a human-in-the-loop decision making processes involving hypothesis suggestion and evaluation. LLMs have also been
exploited in large scale smart literature searches \cite{lee2026llmb}, leveraging knowledge already available in the scientific literature.

The evolutionary algorithm approach taken in this study differs from the above mentioned approaches in that it introduces an explicit population 
of candidate solutions that is manipulated and evaluated through agents and tools. Thus, we do not design a single agent, but rather a 
set of meta rules that instantiate and control agentic operations within a genetic algorithm. In essence, the LLM is treated as one component 
of an optimization algorithm. Similar in philosophy, Caldas et al. \cite{caldas2023bayesian} exploits in-context learning to use the LLM as a surrogate function.
Evolutionary search \cite{abhyankar2026llema} has been successfully applied to materials discovery in the context of wide bandgap semiconductors, 
solid state electrolytes, and insulating dielectrics, where materials are described through crystallographic representations. In the present work, 
we bring these capabilities to the search for synthesizable molecules and represent candidates through synthetic routes.

\begin{figure}[!t]
\centering
\includegraphics[width=0.5\linewidth]{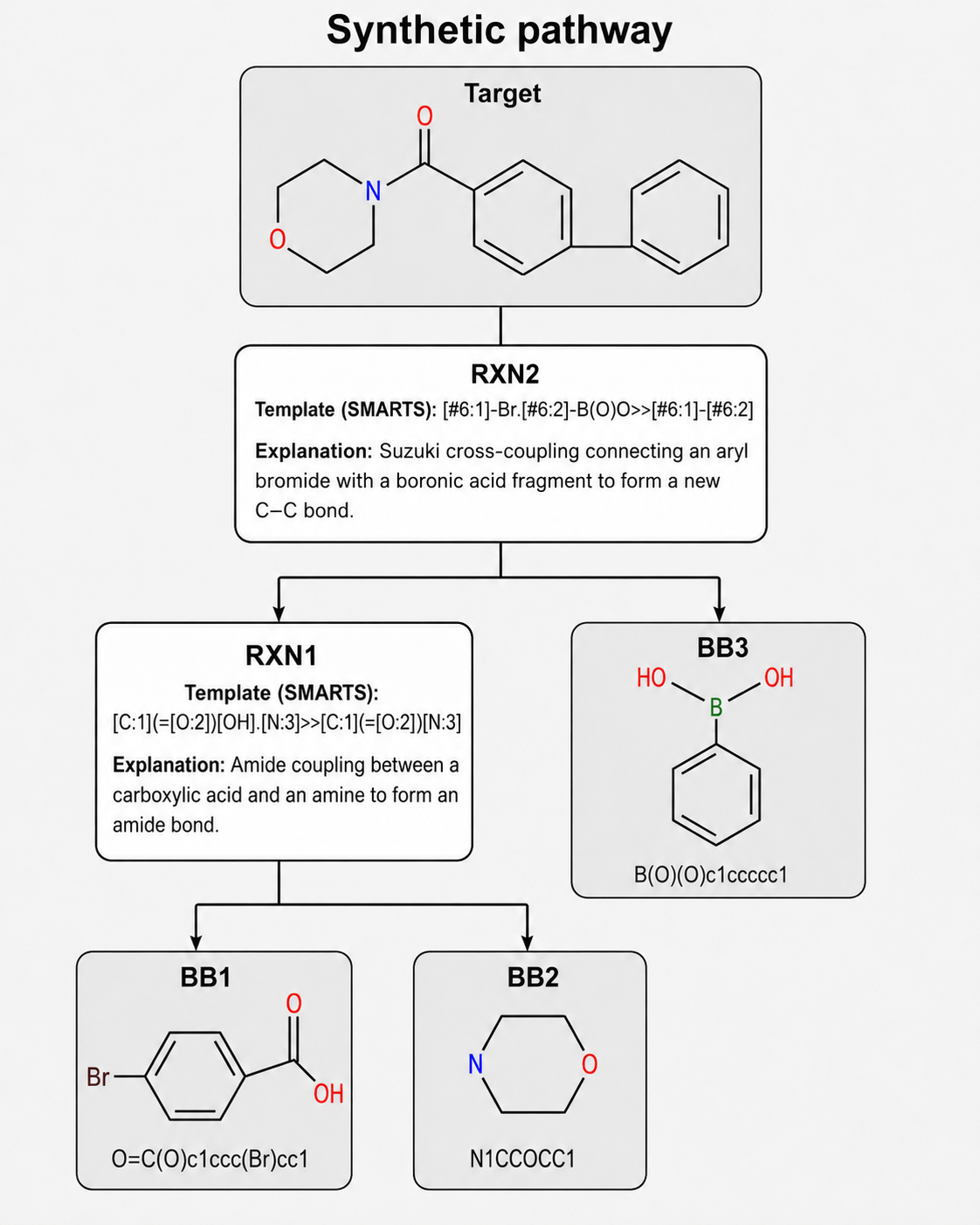}
\caption{Example of an executable synthetic reaction route used as the search object in route native molecular optimization.}
\label{fig:example_reaction_route}
\end{figure}

Our approach differs from synthesis aware generative models and posthoc projection methods \cite{lee2026exploring, gao2024generative} 
in the object that is searched. 
Rather than first proposing molecular structures and subsequently attempting to map them onto feasible chemistry, we formulate 
molecular design as a genetic algorithm over executable synthetic routes. As illustrated in 
Figure~\ref{fig:example_reaction_route}, each candidate is not merely a molecule, but a concrete synthesis program: a sequence 
or tree of reaction steps that combines purchasable building blocks through reaction templates to produce a final product. 
The population of the genetic algorithm is therefore a population of synthetic routes. Each route acts as the genotype, while 
the molecule obtained by executing the route acts as the phenotype on which molecular properties are evaluated. In this 
representation, synthesizability is not introduced as an external penalty or correction, but is imposed by the search space itself.

Within this framework, the role of the LLM is deliberately restricted. It does not act as an unconstrained molecule generator; 
instead, it serves as a chemistry aware mutation operator acting on route genomes. Given parent routes sampled from the current 
population, the LLM proposes local or structural edits, such as building block substitutions, reaction template changes, route 
extensions, or recombinations of useful route fragments from previous candidates. These proposed offspring are then parsed into 
executable route objects, validated against the reaction library, executed with deterministic chemistry tools, and scored by 
downstream molecular oracles. The evolutionary loop therefore maintains explicit control over population update, selection, and 
fitness evaluation, while the LLM contributes chemical priors that guide exploration of the synthesizable route space.

\section{Results}

\subsection{Overall Workflow}

\begin{figure}[!t]
\centering
\includegraphics[width=\linewidth]{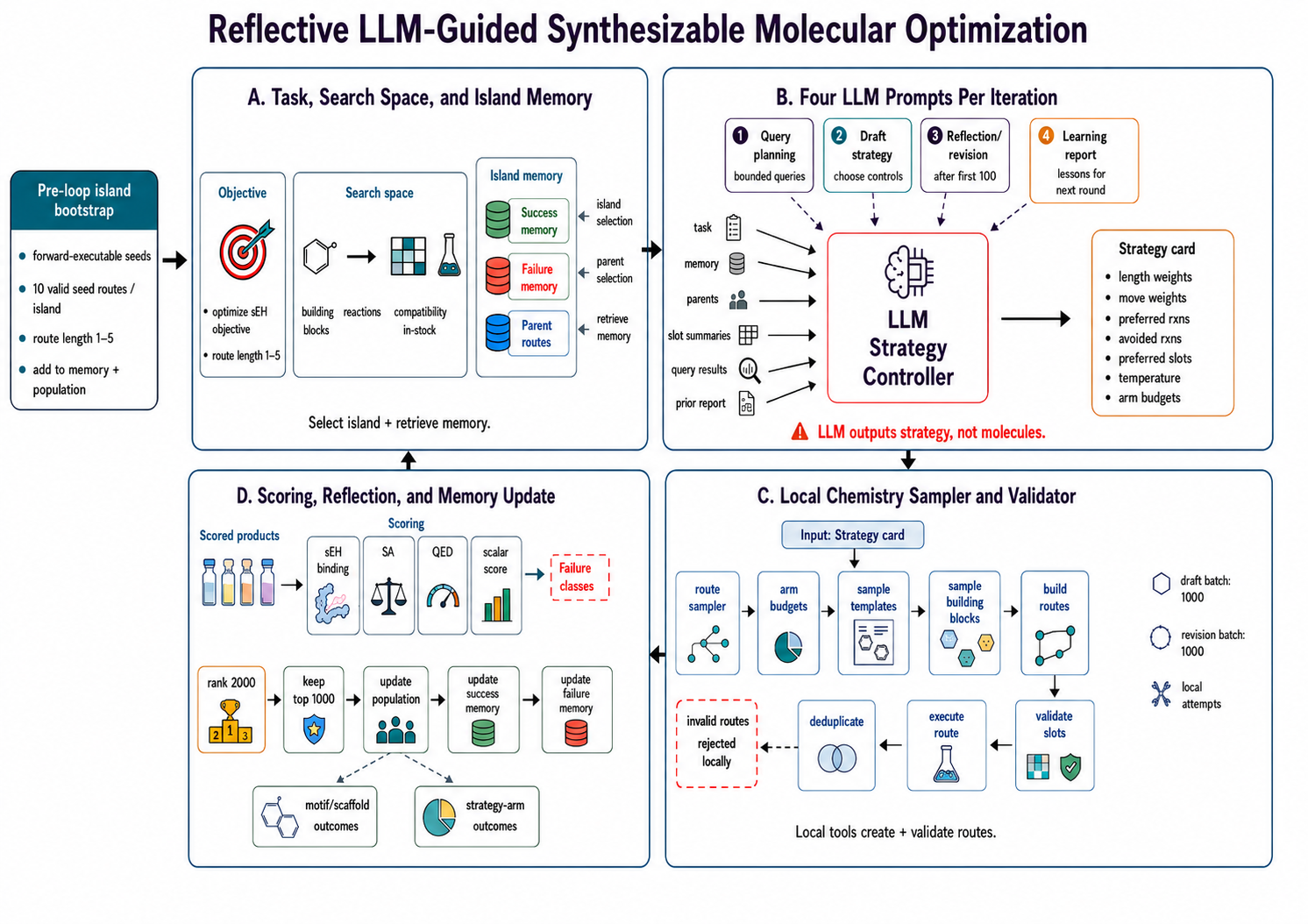}
\caption{System architecture for {\scshape My
Chemical Harness}, showing the interaction between LLM guided strategy generation, route sampling, deterministic execution, scoring, and memory conditioned evolutionary search.}
\label{fig:mch_system_architecture}
\end{figure}

The overall workflow of {\scshape My
Chemical Harness} is schematically illustrated in
Fig.~\ref{fig:mch_system_architecture}. A detailed description of the
methodology and algorithmic components is provided in
Section~\ref{sec:method}. In brief, we formulate synthesizable molecular design
as an evolutionary search problem over synthetic routes. Each individual in the
population corresponds to a candidate synthetic pathway, and the product
molecule obtained by executing that route is evaluated by one or more objective
functions subject to task specific constraints. The population is then
iteratively refined using genetic algorithm principles, including selection,
variation, and fitness based survival
\cite{lange2024large,holland1992genetic}.

To instantiate the search, the system requires a library of purchasable or
otherwise accessible building blocks, a set of reaction templates, and a
compatibility matrix specifying which building blocks can occupy each reaction
slot. The search is organized into population islands. Each island stores its
own successful routes, failed or low scoring routes, parent candidates, and
island local sampling preferences. Before the first LLM guided iteration, islands are seeded 
with executable bootstrap routes, so that the initial prompt contains evaluated chemical examples 
rather than only a global search space description.
In the sEH experiments, bootstrap routes are generated locally through forward execution. The sampler incrementally grows routes with lengths between one and five, retaining only partial routes that can be executed. The resulting products are then scored, and valid seeds are added to both the island memory and the population state.

The workflow can be viewed as four coupled components:
\begin{itemize}
    \item \textbf{Task, search space, and island memory.} The task defines the
    optimization objective and constraints, while each island maintains a compact
    history of successful, failed, and parent routes within the accessible
    synthetic action space.

    \item \textbf{LLM strategy controller.} The LLM acts as a search policy
    controller rather than a molecule generator, using the task, island context,
    and bounded chemistry space tools to assign soft preferences over route
    length, move type, reaction families, reaction slots, motifs, and
    exploration pressure. Unless stated otherwise, all LLM-guided experiments were run using \texttt{deep-seek-v4-flash} as the underlying LLM provider.

    \item \textbf{Local chemistry tools.} Deterministic chemistry code converts
    the strategy into executable candidate routes, enforces reaction slot
    compatibility, removes duplicates, and scores the resulting product
    molecules.

    \item \textbf{Reflection, selection, and memory update.} Scored candidates
    are summarized, ranked, and used to update the island population and memory;
    in reflective runs, these summaries also guide a mid iteration strategy
    revision and the learning context for the next iteration.
\end{itemize}

\paragraph{One reflective optimization iteration.}
In the reflective configuration considered here, each optimization iteration is
a fixed loop in which the LLM plans and revises the search strategy, while
local chemistry code generates and evaluates the molecules. The LLM facing part
of one iteration proceeds as follows:
\begin{enumerate}
    \item \textbf{Local context assembly.} First, the system gathers the
    information that the LLM will be allowed to see from the active island.
    Local code selects the island, retrieves parent routes and memory, and
    summarizes the current reaction and building block action window.

    \item \textbf{LLM query planning.} Next, the LLM asks for a few extra
    resource lookups before deciding how to steer the search. It returns a small
    set of permitted chemistry space tool queries, which local code executes and
    summarizes for the next prompt.

    \item \textbf{LLM draft strategy.} The LLM then proposes an initial plan for
    where the search should spend its effort. Using the task, memory, parent
    routes, query results, and previous learning report, it produces a
    structured search policy rather than explicit molecules.

    \item \textbf{Local first pass execution.} The chemistry engine tests the
    LLM's first plan by turning it into real route proposals. It enforces
    synthetic validity, removes duplicate routes, scores the resulting products,
    and summarizes the first pass outcomes.

    \item \textbf{LLM reflective revision.} The LLM is shown what happened and
    is asked to improve its plan before the iteration finishes. It uses the
    first pass outcome summary to revise the strategy for the remaining
    candidate budget.

    \item \textbf{Local second pass execution and selection.} The revised plan
    is tested in the same chemistry pipeline, and the best candidates from both
    passes are kept. The pooled candidates are ranked by the task score, then
    committed to the island population, archive, and memory.

    \item \textbf{LLM learning report.} Finally, the LLM writes down the lesson
    from the iteration so that the next iteration does not start from scratch. A
    compact report of useful strategies, reaction patterns, motifs, and failure
    modes is carried forward to guide the next iteration.
\end{enumerate}
Thus, the LLM remains a planner and reviewer, while route construction,
validation, scoring, deduplication, and population updates remain deterministic
local operations.

\subsection{sEH as a model system for reflective route optimization}

We first used soluble epoxide hydrolase (sEH) as a controlled molecular design
testbed to study how the reflective route evolution algorithm behaves when the
optimization target is a single protein binding proxy embedded in a
synthesizable chemical space. sEH is a pharmacologically relevant enzyme that
hydrolyzes endogenous epoxy fatty acid mediators, and it has become a common
benchmark for goal directed molecular generation because a pretrained proxy
model for predicted binding is available from the GFlowNet molecular synthesis
setting of Bengio et al.~\cite{bengio2021flow}. Following related
synthesis constrained studies from Gao et al., Cretu et al., and the NVIDIA
ReaSyn work~\cite{gao2024generative,cretu2024synflownet,lee2026exploring}, we
evaluate not only the sEH binding score but also practical molecular quality and
synthesis metrics. Synthetic accessibility (SA) measures how difficult a
molecule is expected to be to synthesize, with lower raw SA values indicating
easier synthesis. QED measures drug likeness on a normalized scale and
AiZynthFinder reports whether a retrosynthesis planner can identify a feasible
route from purchasable building blocks.

\begin{table}[!b]
\centering
\caption{\textbf{Synthesizable goal directed molecular optimization results on the sEH proxy, with baseline results taken from the NVIDIA Reasyn study.}
Baseline values are reproduced from Lee et al.~\cite{lee2026exploring} and
reported as the means and standard deviations of 3 runs. Our result reports the
final cumulative top-1000 mean across the three reflective LLM guided runs
analyzed in this work. The best results are highlighted in bold.}\vspace{5pt}
\label{tab:seh_proxy_results}
\resizebox{0.92\textwidth}{!}{%
\begin{tabular}{lcccc}
\hline
Method & sEH & SA $\downarrow$ & QED & AiZynth. \\
\hline
SyntheMol & $0.64 \pm 0.01$ & $3.08 \pm 0.01$ & $0.63 \pm 0.01$ & 0.82 \\
SynFlowNet & $0.92 \pm 0.01$ & $2.92 \pm 0.01$ & $0.59 \pm 0.02$ & 0.65 \\
Reasyn & $0.96 \pm 0.00$ & $2.05 \pm 0.01$ & $\mathbf{0.75 \pm 0.01}$ & $0.97 \pm 0.01$ \\
Reflective LLM (ours) & $\mathbf{0.984 \pm 0.005}$ & $\mathbf{1.996 \pm 0.027}$ & $0.692 \pm 0.015$ & $\mathbf{0.994 \pm 0.004}$ \\
\hline
\end{tabular}%
}
\end{table}

For the sEH case study, we treated LLM control as an ablation variable and used three controller labels. In the \textit{deterministic} setting, no LLM calls are made. Instead, local code samples routes from fixed or hand-specified search preferences, executes and scores them, and updates the population using the same deterministic chemistry and selection rules applied across all runs. In the \textit{non-reflective} LLM setting, the LLM is called once per iteration to choose a strategy for local route sampling, but the reflective part of the workflow is omitted. As a result, the LLM does not observe same-iteration outcomes, revise its strategy, or write a learning report before the population update. In the \textit{reflective} LLM setting, the full agentic workflow is used. After an initial local sampling pass, the LLM receives a summary of the successes, failures, and motifs observed in that pass, revises its strategy, and the local sampler uses the remaining candidate budget under the revised strategy.

To isolate the effect of reflective LLM agent from the underlying local
chemistry engine, we compared it to the non-reflective LLM
control, and the deterministic strategy controller with the same route execution,
scoring, deduplication, and population update machinery. The reflective
agent evaluated 1000 locally sampled candidates before reflection and
1000 after reflection in each iteration. The deterministic control therefore
evaluated a total of 2000 candidates per iteration to match the local candidate budget.
Table~\ref{tab:seh_proxy_results} compares the final reflective LLM population
with synthesis constrained baselines from the ReaSyn study. The reflective LLM
run achieves the highest cumulative top-1000 mean sEH score
($0.984 \pm 0.005$), the lowest raw SA score ($1.996 \pm 0.027$), and the
highest AiZynthFinder success rate ($0.994 \pm 0.004$), while its QED
($0.692 \pm 0.015$) remains below the ReaSyn value. Thus, in this setting the
reflective controller improved the binding and synthesis facing metrics most
directly targeted by the executable search, without asking the LLM to design
the molecules directly.

\begin{figure}[!t]
\centering
\includegraphics[width=\linewidth]{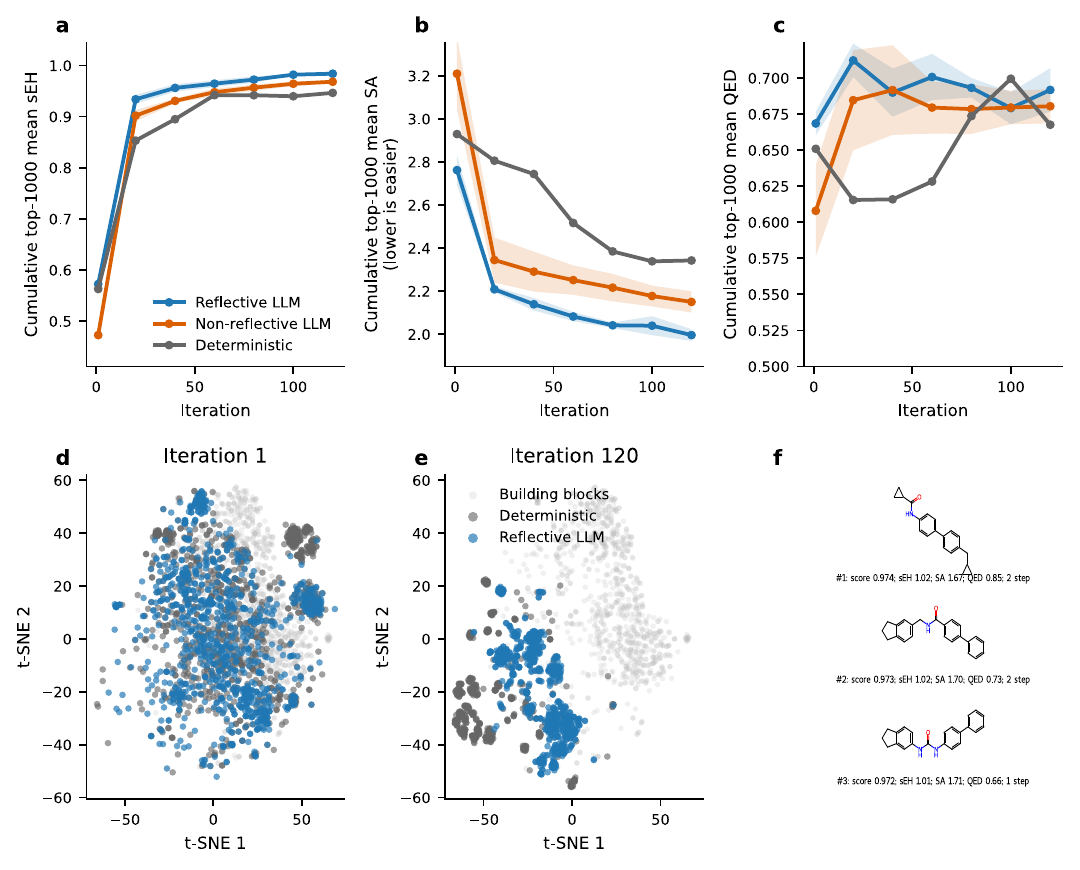}
\caption{\textbf{sEH case study for reflective route optimization.}
(a) Cumulative top-1000 mean normalized sEH score over 120 iterations for
the reflective LLM agent, the non-reflective LLM, and the deterministic controller. Shaded
regions show the standard deviation across three independent LLM runs; the
deterministic control is a single run. (b,c) The same cumulative top-1000
products summarized by raw SA score (lower is easier) and QED. (d,e) Shared
t-SNE projection of Morgan fingerprints using Tanimoto distances for the
selected reflective and deterministic runs at the first and final iterations,
with 1000 sampled Enamine/Synformer building blocks in gray. (f) Three
highest scoring final products from the reflective LLM runs.}
\label{fig:seh_reflective_case_study}
\end{figure}

Figure~\ref{fig:seh_reflective_case_study} summarizes how this policy level
control changes the search over time. Across 120 iterations, the final
cumulative top-1000 scalar objective was $0.955 \pm 0.002$ for reflective LLM
agent, $0.945 \pm 0.002$ for non-reflective LLM control, and $0.930$ for the
deterministic control. Panels a--c show that the reflective agent gives
the strongest sEH trajectory while preserving easier synthesis and comparable
drug likeness in the same cumulative top-1000 molecule pool. In particular,
the final cumulative top-1000 component means were sEH
$0.984 \pm 0.005$, SA $1.996 \pm 0.027$, and QED $0.692 \pm 0.015$ for the
reflective agent, compared to sEH $0.946$, SA $2.343$, and QED $0.668$
for the deterministic control. Panels d and e show the associated change in
chemical space coverage. At the first iteration, the reflective and
deterministic products largely overlap near the sampled Enamine/Synformer
building block reference set; by the final iteration, the two searches occupy
more separated regions of the shared t-SNE projection. This indicates that the
optimization is not merely selecting the same local products more strongly, but
is opening distinct productive regions of synthesizable chemical space.

The route records provide a simple audit of which search strategy was most
productive. Among the final cumulative top-1000 reflective products across the
three runs, 60.1\% were one step routes and 34.5\% were two step routes, with
only 5.4\% requiring three or more reaction steps. The most common successful
strategy was therefore not a long route elaboration policy, but repeated
generation and refinement of short executable routes, with the LLM steering
which reaction and building block regions the local sampler should emphasize.
The chemical products in panel f are consistent with this interpretation. The
three highest scoring reflective products are compact aromatic amides or ureas,
and among the final top-10 reflective products all contained aromatic rings,
five contained amide motifs, and four contained urea motifs. This behavior is
chemically plausible: crystallographic and medicinal chemistry studies of sEH
inhibitors show that urea like hydrogen bonding motifs and hydrophobic or aryl
substituents are common ways to engage the catalytic pocket and enzyme
tunnel~\cite{argiriadi2000binding,gomez2006human,kim2007disubstituted,huang2010piperazino,lee2014optimized}.
The reflective policy therefore enriched interpretable sEH relevant chemistry
while preserving the auditability of an executable, locally validated route
generator.

\subsection{TDC oracle benchmarks}

We next asked whether the same reflective, synthesis constrained search
procedure could be applied beyond the sEH proxy to the 13 molecular design
oracles commonly used in the Therapeutics Data Commons (TDC) goal directed
generation benchmark~\cite{huang2021tdc}. This oracle suite, evaluated with
area under the top-10 curve in the practical molecular optimization benchmark of
Gao et al.~\cite{gao2022sample}, has also been used in recent
synthesis constrained molecular generation studies from Lee et al.~\cite{lee2026exploring} and Sun et
al.~\cite{sun2025procedural}. These objectives include
single target activity oracles and multiparameter optimization tasks, so each
oracle defines a separate optimization problem with a different notion of
molecular quality. We therefore ran each TDC objective independently, using the
same route execution and reflective strategy control machinery as in the sEH
experiments. Table \ref{tab:tdc_oracle_results} compares our results with the
synthesis based methods reported in the NVIDIA Reasyn study.

\begin{table}[!t]
\centering
\caption{\textbf{Synthesizable goal directed molecular optimization results on the TDC oracles, with baseline results taken from the NVIDIA Reasyn study.}
All results are AUC top-10. The Reasyn, SynthesisNet, and SynNet values are taken from Lee et al.~\cite{lee2026exploring}.
The best value among the displayed methods is highlighted in bold.}
\vspace{5pt}
\label{tab:tdc_oracle_results}
\resizebox{\textwidth}{!}{%
\begin{tabular}{lcccc}
\hline
Method & Reasyn & SynthesisNet & SynNet & Reflective LLM (ours) \\
\hline
amlodipine\_mpo & 0.620 & 0.608 & 0.567 & \textbf{0.6436} \\
celecoxib\_rediscovery & \textbf{0.810} & 0.582 & 0.443 & 0.5362 \\
drd2 & \textbf{0.977} & 0.960 & 0.969 & 0.9752 \\
fexofenadine\_mpo & 0.788 & 0.791 & 0.764 & \textbf{0.7934} \\
gsk3b & \textbf{0.889} & 0.848 & 0.790 & 0.8074 \\
jnk3 & \textbf{0.695} & 0.639 & 0.631 & 0.5228 \\
median1 & 0.274 & \textbf{0.305} & 0.219 & 0.2415 \\
median2 & \textbf{0.259} & 0.257 & 0.237 & 0.2415 \\
osimertinib\_mpo & \textbf{0.823} & 0.810 & 0.797 & 0.8076 \\
perindopril\_mpo & \textbf{0.561} & 0.524 & 0.559 & 0.5585 \\
ranolazine\_mpo & 0.752 & 0.741 & 0.743 & \textbf{0.7660} \\
sitagliptin\_mpo & 0.314 & 0.313 & 0.026 & \textbf{0.4522} \\
zaleplon\_mpo & 0.460 & \textbf{0.528} & 0.341 & 0.5076 \\
\hline
Average score & \textbf{0.633} & 0.608 & 0.545 & 0.6041 \\
\hline
\end{tabular}%
}
\end{table}

\section{Discussion and Conclusions}

This work sits within a rapidly growing set of approaches for designing
synthesizable molecules with machine learning and language models. SyntheMol and
related work from Gao et al.\cite{gao2024generative} showed that generative models can be constrained
toward synthetically accessible chemical space; ReaSyn from Lee et al.\cite{lee2026exploring} and
procedural synthesis from Sun et al.\cite{sun2025procedural} further emphasized that molecular
generation should be coupled to explicit synthetic pathways rather than treated
as structure generation followed by a synthesis check. SynLlama\cite{sun2025synllama} and MolReAct\cite{li2026molreact} move closer to language-model-driven molecular design by using LLMs to generate synthesizable analogues or to guide action spaces for reinforcement learning over validated reaction templates.
Our contribution is complementary. We use the LLM neither as an unconstrained
SMILES generator nor as the component that validates chemistry. Instead, the LLM
acts as a strategy controller inside a route-native evolutionary search, while
local deterministic code constructs, executes, validates, deduplicates, and
scores the candidate routes.

A central lesson from our experiments is that the interface to the LLM is itself part of the algorithm. The relevant chemical space is too large to expose directly to the model. In our setting, hundreds of thousands of building blocks and many reaction templates cannot simply be pasted into a prompt. The useful question is therefore not whether an LLM can understand the entire search space at once, but how that space should be partitioned, summarized, and refreshed so that the model can make productive high-level decisions. Most of the mechanical work should remain local, including the sampling of compatible reactants, the validation of reaction slots, the execution of RDKit reactions, the rejection of invalid pathways, and the evaluation of objective functions. The LLM is most useful when it receives compact evidence about what has worked and failed, and then uses that evidence to bias the next local search toward promising reaction families, motifs, move types, or route lengths.

This separation also gives the method an important form of auditability and
trust. Because accepted molecules are produced by RDKit execution and validated
by local deterministic code, the algorithm does not depend on the LLM discovering
chemically valid products. In the worst case, an unhelpful LLM strategy may fail
to improve over deterministic search, but it cannot by itself introduce a
hallucinated molecule or unsupported synthetic step into the accepted population.
Every retained candidate is tied to an executable route under the available
template library, and every score is computed by the configured local oracle.
For chemists, this assigns language models a more reliable role, allowing them to guide where the search should go while leaving the chemistry engine to determine which molecules and synthetic steps are actually valid.

The sEH experiments illustrate both the promise and the limitation of this
division of labor. Reflection did not transform the optimization problem, and a
strong deterministic sampler remained a competitive baseline. However, the
reflective controller did change the allocation of search effort and enriched
chemically interpretable motifs, including urea- and amide-containing products
consistent with known sEH inhibitor chemistry. Thus, the main value of the LLM
was adaptive prioritization over a constrained synthetic action space. The TDC experiments reinforce the same point by showing that different oracles reward different molecular features, so future versions will likely require task-specific memory, prompt summaries, and action-window construction rather than a single universal prompting recipe.

More broadly, {\scshape My Chemical Harness} belongs to a larger movement toward AI systems
that assist scientific discovery by coupling language, tools, search, and
experimental feedback. Recent multi-agent systems such as Co-Scientist and Robin
show how LLM agents can generate hypotheses, critique them, and connect them to
experimental validation in biomedical settings
\cite{gottweis2026coscientist,ghareeb2026multiagent}. In chemistry, Co-Scientist
demonstrated that language-model agents can plan and execute parts of
experimental workflows when connected to appropriate tools and automation
\cite{boiko2023autonomous}. Our work addresses an upstream piece of the same
future pipeline: generating synthesizable molecular hypotheses and
their routes before committing resources to synthesis and assay. The next
challenge is to connect route-native algorithms such as this one to real
closed-loop medicinal chemistry, including reagent availability, reaction
condition selection, purification, characterization, biological testing, and
feedback from failed experiments.

In conclusion, LLMs can accelerate molecular discovery when their role is carefully constrained. The goal is not to ask the model to replace chemistry, but to place it within a harness in which local tools enforce chemical reality. In this setting, the LLM acts as a high-level search partner, improving exploration when useful, remaining harmless when it is not, and staying bounded at every step by executable synthetic chemistry.

\section{Methods}
\label{sec:method}

We formulate {\scshape My
Chemical Harness} as an LLM guided evolutionary search procedure over executable
synthetic routes. It combines a task specification, a bounded route defined search space, an LLM 
strategy controller, deterministic route execution, oracle scoring, and island local memory. This
organization is related to AlphaEvolve \cite{novikov2025alphaevolve} LLM guided evolutionary 
search, but differs in the object manipulated by the LLM. Rather than proposing program hypotheses 
directly, the LLM controls a set of variation operators that act as tools for modifying a population 
of synthetic routes. The population therefore consists of executable synthesis routes, and the fitness
of a route is obtained only after local route execution produces a final molecule.

\subsection{Route and pathway representation}
\label{subsec:route_representation_revised}

The search space is defined by a set of building blocks $\mathcal{B}$, a set of
reaction templates $\mathcal{R}$, and a compatibility matrix $\mathcal{M}$. Each
building block in $\mathcal{B}$ is represented chemically by a SMILES string,
where SMILES
denotes the Simplified Molecular Input Line Entry System. SMILES strings
describe concrete molecular graphs, for example
\texttt{NS(=O)(=O)NCC(=O)O}. A reaction template in $\mathcal{R}$ is represented
by SMARTS, where SMARTS denotes SMILES Arbitrary Target Specification. SMARTS is
a pattern language for SMILES reactants and products; for example, an
amide forming template matches an amine and a carboxyl like reactant pattern and
rewrites them as an amide product pattern, as in
\texttt{[C:1](=[O:2])[OH].[N:3]>>[C:1](=[O:2])[N:3]}.
The functional chemistry is the molecular graph encoded by SMILES and the
reaction pattern encoded by SMARTS.

We distinguish reaction templates from actual reaction steps. A template
$R\in\mathcal{R}$ is an abstract SMARTS rule. An instantiated reaction step
$s\in\mathcal{S}$ is obtained only after concrete reactant SMILES are supplied to
that rule and RDKit applies the SMARTS transform to produce product SMILES.
Thus, $\mathcal{S}$ is the set of executable reaction steps induced by applying
templates in $\mathcal{R}$ to compatible reactants from $\mathcal{B}$ or from
earlier intermediates, subject to $\mathcal{M}$. The template describes what kind
of transformation is possible, while the instantiated step records that the
transformation was applied to specific molecules.

A route is the machine executable candidate object. In the implementation, it is
a list of reaction steps:
\begin{equation}
    r=(s_1,\ldots,s_L), \qquad s_i\in\mathcal{S}.
\end{equation}
Each step $s_i$ is formed by choosing a template from $\mathcal{R}$ and supplying
reactants from $\mathcal{B}$ or from products of earlier steps. Local execution
resolves these references to reactant SMILES, applies the corresponding SMARTS
transform, and stores the product SMILES. A pathway is the chemical synthesis
tree induced by the same route object: building blocks are leaves, reaction
steps are internal transformations, and the final executed step is the root
product. Thus,
\emph{route} refers to the executable data structure, while \emph{pathway}
refers to the chemical interpretation of that structure.

The compatibility matrix $\mathcal{M}$ specifies which building blocks can occupy
each reactant slot of each reaction template. It therefore acts as a discrete filter
on the route search space: the sampler proposes only template  reactant
combinations that are allowed by this matrix before RDKit verifies the full
SMARTS match during execution.

This representation makes synthesizability structural rather than post hoc. A
candidate is accepted for scoring only if each selected template resolves to a
stored SMARTS rule, every reactant reference resolves to a building block SMILES
or a previous intermediate product, the reactants match the required SMARTS
patterns, and RDKit execution yields a valid product at each step. Route
signatures are stored for reproducibility and deduplication, but they are not
chemical representations. Product level duplicates are tracked separately,
because two different routes can lead to the same canonical product SMILES.

\subsection{Problem formulation}
\label{subsec:problem_formulation}

Let $\mathcal{X}_{\mathrm{syn}}(\mathcal{B},\mathcal{R},\mathcal{M})$ denote the
route defined synthesizable search space induced by the building blocks,
reaction templates, and compatibility matrix. Each candidate
$r\in\mathcal{X}_{\mathrm{syn}}$ is an executable synthetic route, and we write
$\mathbf{m}(r)$ for the product molecule obtained by executing route $r$. If
execution fails, $\mathbf{m}(r)=\bot$.
If the discovery task had only one molecular property of interest, the route
optimization problem would be
\begin{equation}
    r^\star =
    \argmax_{r\in\mathcal{X}_{\mathrm{syn}}}
    f(\mathbf{m}(r)),
\end{equation}
where $f$ is a black box property oracle evaluated on the executed product. In
practice, molecular design rarely reduces to a single property. A candidate must
satisfy route level and molecular constraints
\begin{equation}
    C=\{c_1,c_2,\ldots,c_k\}
\end{equation}
while jointly optimizing several objectives $f_1,\ldots,f_n$. Each constraint
$c_j$ corresponds to a property or validity function $g_j$ and takes one of the
canonical forms
\begin{equation}
    c_j: g_j(\mathbf{m}(r))\in[l_j,u_j]
    \quad \mathrm{or} \quad
    c_j: g_j(\mathbf{m}(r))\geq l_j
    \quad \mathrm{or} \quad
    c_j: g_j(\mathbf{m}(r))\leq u_j,
\end{equation}
where $l_j,u_j\in\mathbb{R}$ are task specific thresholds. Route validity is also
a hard constraint: if $\mathbf{m}(r)=\bot$, the candidate is treated as an
invalid genotype rather than as a low scoring molecule.

The constrained multi objective task is therefore to identify executable routes
whose products satisfy all constraints while achieving favorable trade offs
across objective functions. A scalarized version of this task is
\begin{equation}
    r^\star =
    \argmax_{r\in\mathcal{X}_{\mathrm{syn}},\ r\models C}
    \sum_{i=1}^{n} w_i f_i(\mathbf{m}(r)),
\end{equation}
where $w_i$ is the weight of objective $i$ and $r\models C$ denotes that the
route executes and satisfies all hard constraints. This is the score used for
scalar selection runs and for logging.  For the sEH experiments, $T$ asks the system to 
discover routes of one to fivesteps whose products have strong soluble epoxide hydrolase 
binding while maintaining practical synthetic accessibility and drug like molecular quality.
The main objectives are sEH binding, synthetic accessibility, and QED. These
objective values are not produced by the LLM. They are computed by the local
scoring pipeline after the final product SMILES has been generated by route
execution. The detailed normalization used to convert raw objective values into
scalar rewards is provided in Appendix~\ref{app:objective_scoring}.

\subsection{LLM guided evolutionary algorithm}
\label{subsec:llm_guided_ea_revised}

The algorithm is a genetic algorithm in the standard sense that it maintains a
population, selects parents, applies variation operators, evaluates fitness, and
updates the population across generations \cite{holland1992genetic}. In this
work, the genotype is a synthetic route and the phenotype is the molecule
obtained by executing that route. The search is organized over a set of islands
$\mathcal{Z}$; each island $z\in\mathcal{Z}$ maintains its own population
$P_{t,z}$ and memory $M_{t,z}$ (too be explained in detail below). 
Selection, survival and mutation are handled by local code, while the LLM supplies 
chemistry aware search guidance, by selecting how to use the tools in charge of the mutations.

At iteration $t$, the system selects an island $z_t$ and builds a prompt
\begin{equation}
    p_t = S_t(T,C,R,A_t,P_{t,z_t},M_{t,z_t},H_t).
    \label{eq:promt}
\end{equation}
The LLM samples a strategy object
\begin{equation}
    \sigma_t \sim \pi_\theta(\cdot \mid p_t),
\end{equation}
rather than a molecule or a full route. The strategy object is a compact JSON
card specifying how the local sampler should allocate search effort. It can
include route length preferences, move type preferences, preferred or avoided
reaction templates, preferred reaction slots, preferred building blocks,
exploration temperature, strategy arms, and requests for future action window
coverage.
Strategy arms are budgeted sub strategies within $\sigma_t$: each arm can
emphasize a different local search mode, such as exploitation of successful
parents or exploration of underused reaction families. Their concrete
interpretation as operator controllers is described in
Section~\ref{subsec:variation_operators_revised}.

The local sampler then defines the actual offspring proposal:
\begin{equation}
    Q_t \sim K_{\sigma_t}(\cdot \mid A_t,P_{t,z_t},M_{t,z_t},H_t).
\end{equation}
This separation is central to the harness. The LLM chooses search pressure; the
local sampler creates concrete routes, validates reaction compatibility,
executes each route, deduplicates candidates, and calls the scorer.

The terms in $p_t$ define the prompt contract for the route native strategy
controller. We now proceed to specify all the elements of the promt Eq. \ref{eq:promt}.
\begin{itemize}
    \item $T$ is the task specification: objectives, score directions, route
    limits, and constraints. In the reflective sEH runs, this appears near the
    top of the prompt as a compact objective block. It tells the model that
    optimization is multiobjective, but that objective values will be computed
    after local route execution.
\begin{lstlisting}[style=mchprompt]
Task:
- task_name: synthesizable_seh
- description: Discover executable, synthesizable molecules with strong soluble
  epoxide hydrolase binding while preserving practical synthesis and drug-like
  molecular quality.
- objectives: {
    "seh_binding_score": {"direction": "maximize", "lower": 0.0, "upper": 1.0},
    "synthetic_accessibility": {"direction": "maximize", "lower": 0.0, "upper": 1.0},
    "qed_drug_likeness": {"direction": "maximize", "lower": 0.0, "upper": 1.0}
  }
Route step range: 1 to 5
\end{lstlisting}

    \item $C$ is the set of hard constraints enforced by local chemistry tools.
    These constraints are not left for the LLM to decide. A proposed preference
    or sampled route must refer to known reaction templates, known building
    blocks, valid intermediate references, reaction slots allowed by
    $\mathcal{M}$, and an RDKit executable transformation. A route level
    constraint is therefore a
    chemistry constraint over resolved SMARTS and SMILES, with handles used only
    to retrieve the corresponding records:
\begin{lstlisting}[style=mchprompt]
local validation example:
  template: synformer_template_009
  SMARTS: [NX3;!$(N[C]=[S,O,N]);$(N[#6,S]);!$(N[#7]);H1,H2:1].
          [#6:4][C:5](=[O:6])[OH,O-]>>[N:1][C:5](=[O:6])[#6:4]
  reactants:
    - step_3
    - enamine_0059419 [SMILES: NS(=O)(=O)NCC(=O)O]
  checks:
    template resolves to SMARTS; building block resolves to SMILES;
    valid step_3 intermediate; slot-compatible reactants;
    RDKit execution yields a product;
    route signature has not already been accepted.
\end{lstlisting}

    \item $R$ is the rule set shown to the LLM, including JSON output rules and
    the restriction that only listed chemistry may be selected. The prompt gives
    the LLM SMARTS and SMILES for reasoning, but the output contract remains a
    compact JSON strategy over listed fields, because local code must resolve the
    requested reaction templates and building blocks.
\begin{lstlisting}[style=mchprompt]
Strategy rules:
1. You do not write concrete synthesis routes.
2. Local code samples compatible building blocks and executes RDKit reactions.
3. Choose only reaction-template labels shown in the available SMARTS families.
4. preferred_reaction_slots and request_more_slots must use available templates
   and valid integer slot indices.
5. Do not invent templates, building blocks, SMILES, SMARTS, or JSON fields.
Return only JSON. No markdown. No prose outside JSON.
\end{lstlisting}

    \item $A_t$ is the current action window: sampled reaction templates,
    compatible building blocks, slot summaries, and optional chemistry space
    query results. It is the finite menu of chemistry that the local sampler may
    draw from during iteration $t$. Quantities such as route length weights,
    move type weights, and route position weights are not elements of $A_t$; they
    are controls in the strategy object $\sigma_t$ that bias how the sampler uses
    this menu. In the referenced reflective configuration, the prompt window is
    built from 60 reaction templates and 20000 slot balanced building blocks, with
    each template label followed by its SMARTS and slot level compatibility
    counts:
\begin{lstlisting}[style=mchprompt]
Available reaction-slot families this iteration:
- synformer_template_009
  SMARTS: [NX3;!$(N[C]=[S,O,N]);$(N[#6,S]);!$(N[#7]);H1,H2:1].
          [#6:4][C:5](=[O:6])[OH,O-]>>[N:1][C:5](=[O:6])[#6:4]
  name: amide-forming C-N acylation template
  class: amide formation / C-N acylation
  features: carbonyl, C-N bond formation, sulfur chemistry
  reactants: 2
  slot 0 role: acyl donor or carboxyl-like reactant; compatible_options=7341
  slot 1 role: amine-like nucleophile; compatible_options=3001
\end{lstlisting}
    Optional chemistry space queries further expand $A_t$ with concrete
    compatible examples, still tied to the same finite local catalog:
\begin{lstlisting}[style=mchprompt]
compatible examples for synformer_template_009 slot 1:
- enamine_0059419 [SMILES: NS(=O)(=O)NCC(=O)O]
  motifs: carboxylic acid, sulfonamide, amine
- enamine_0028930 [SMILES: Nc1cccc2c1C(=O)NC2=O]
  motifs: amide, aniline-like amine
\end{lstlisting}

    \item $P_{t,z_t}$ is the active island population from which parent routes
    are selected. Parents are shown as route trees or compact local chemistry
    summaries. They include the SMARTS of each reaction step, the reactant SMILES
    when the reactant is a catalog building block, and \texttt{step\_i}
    references when the reactant is an intermediate:
\begin{lstlisting}[style=mchprompt]
Parent routes (compact local chemistry summary):
- parent_1: candidate_id=bootstrap_forward_executable_routes_0_0004
  route_length: 3
  step_1: synformer_template_057
    SMARTS: [#6;!$(C=[O,S,N]):1][OH,SH:2].
            [Cl,Br,I][#6;!$(C=[N,O,S]):3]>>[#6:1][O,S:2][#6:3]
    reactants:
      enamine_0050400 [SMILES: CC(C#N)C(O)c1ccc(Cl)cc1]
      enamine_0202221 [SMILES: FC(F)(Br)n1cnc(I)c1]
  step_2: synformer_template_038
    SMARTS: [#6;!$(C=[S,O,N]):1][Cl,$(OS(=O)(=O)[#6;!R])].
            [#6;!$(C=[S,O,N]):2][SH:3]>>[#6:1][S:3](=O)(=O)[#6:2]
    reactants: step_1; enamine_0022922 [SMILES: CC1CCN(c2nnc(S)n2CC2CCCO2)CC1]
  step_3: synformer_template_084
    SMARTS: [#6:1][C:2]#[N:3]>>[#6:1][c:2]1[n:3][nH][n][n]1
    reactants: step_2
\end{lstlisting}

    \item $M_{t,z_t}$ is the island memory, containing successes, failures, and
    low scoring but valid examples. During iteration $t$, newly accepted
    candidates are collected in $V_t$ as route product score triples
    $(r,x,\mathbf{y})$, and failed or invalid routes are collected in $F_t$.
    These temporary sets are folded into memory after local execution. Memory
    carries outcome evidence: final product SMILES, score components, motifs,
    scaffold SMILES, route length, and route steps with decoded SMARTS/SMILES.
    This lets the LLM connect a strategy choice to molecular outcomes rather than
    to bookkeeping labels:
\begin{lstlisting}[style=mchprompt]
Success memory:
- success_10: candidate_id=bootstrap_forward_executable_routes_0_0005
  score: 0.297
  score_components: seh=0.152, sa_norm=0.495, qed_norm=0.243,
                    logp=3.326, mw=890.039, tpsa=244.260
  final_smiles: COc1ncc(C=C(C2CCCCC2)C2(CN(Cc3cnc4c(c3)CCN4)
                S(=O)(=O)c3c(C)sc(C)c3C=O)N3C(=O)c4cccc(c4C3=O)
                N2C(=O)CNS(N)(=O)=O)cn1
  motifs: amide, sulfonamide, amine, aniline-like amine, pyridine
  route_steps:
    step_4=synformer_template_009
      SMARTS: [NX3;!$(N[C]=[S,O,N]);$(N[#6,S]);!$(N[#7]);H1,H2:1].
              [#6:4][C:5](=[O:6])[OH,O-]>>[N:1][C:5](=[O:6])[#6:4]
      reactants: step_3; enamine_0059419 [SMILES: NS(=O)(=O)NCC(=O)O]
  local_lesson_hint: explore nearby chemistry around amide, sulfonamide, amine.
\end{lstlisting}

    \item $H_t$ is the historical state used for cross iteration bookkeeping and
    control feedback. It is distinct from the island memory $M_{t,z_t}$:
    $M_{t,z_t}$ is the prompt facing evidence buffer for the currently selected
    island, containing route and molecule examples that help the LLM reason about
    chemistry within that island. By contrast, $H_t$ records run level history
    such as previous strategies, route signatures, rejected routes from $F_t$,
    accepted route summaries from $V_t$, strategy arm outcomes, reaction
    outcomes, and learning reports. This state is partly shown to the LLM in
    reflection and learning report prompts and partly consumed by local code for
    deduplication, slot boosts, island selection, and arm accounting. A
    reflective prompt may show which arms produced valid candidates, which final
    SMILES scored well, and which failure classes should change the next
    strategy:
\begin{lstlisting}[style=mchprompt]
Draft outcome summary:
- strategy_arm: exploit_successful_candidates
  valid_count: 8
  top_scored_candidate:
    final_smiles: CCC(CC)CCCOc1cc2c(cc1C)nc(-c1cccnc1N1CCCC1)n2C
    reaction_templates: [synformer_template_021, synformer_template_057]
    move_type: delete_route_step
    property_values:
      seh_binding_score: 0.539
      synthetic_accessibility: 0.819
      qed_drug_likeness: 0.780
    scaffold_smiles: c1cnc(N2CCCC2)c(-c2nc3ccccc3[nH]2)c1
- near_miss:
    failure_class: too_heavy
    final_smiles: CC1CCN(c2nnc(S(=O)(=O)c3ccc(...))n2CC2CCCO2)CC1
    lesson: reduce high-molecular-weight route extensions.
\end{lstlisting}
\end{itemize}

\begin{algorithm}[H]
\caption{LLM guided route native evolutionary search}
\label{alg:llm_route_native_evolution}
\begin{algorithmic}[1]
\Require $T,C,R,\mathcal{R},\mathcal{B},\mathcal{M},f,\pi_\theta,N,b,\mathcal{Z}$
\Ensure Best valid routes and products, or final Pareto set
\Statex \textcolor{green!50!black}{\(\triangleright\) Initialize}
\ForAll{$z\in\mathcal{Z}$}
    \State Initialize $P_{0,z}\gets\operatorname{InitPop}(\mathcal{R},\mathcal{B},\mathcal{M},C)$.
    \State Initialize $M_{0,z}\gets\operatorname{InitMem}()$ with empty success, failure, and low score memory.
\EndFor
\State Set the historical state to $H_0\gets\emptyset$.
\For{$t=1$ to $N$}
    \Statex \textcolor{green!50!black}{\(\triangleright\) Prompt and strategy}
    \State Select the active island $z_t\gets\operatorname{SelectIsland}(\mathcal{Z},H_t)$.
    \State Build $A_t\gets\operatorname{BuildWindow}(\mathcal{R},\mathcal{B},\mathcal{M},H_t)$.
    \State Build prompt $p_t \gets S_t(T,C,R,A_t,P_{t,z_t},M_{t,z_t},H_t)$.
    \State Initialize the valid candidate set and failed route set, $V_t\gets\emptyset$ and $F_t\gets\emptyset$.
    \Statex \textcolor{green!50!black}{\(\triangleright\) Local route generation}
    \If{reflective mode}
        \State Run Algorithm~\ref{alg:reflection_refinement} to obtain the evaluated batch $B_t$.
        \State Split $B_t$ into valid candidates $V_t$ and failed routes $F_t$.
    \Else
        \State Sample strategy $\sigma_t \sim \pi_\theta(\cdot\mid p_t)$ from the LLM.
        \For{$j=1$ to $b$}
            \State Sample route $r_{t,j}\sim K_{\sigma_t}(\cdot\mid A_t,P_{t,z_t},M_{t,z_t},H_t)$.
            \State Execute the route locally, $x_{t,j}\gets\mathbf{m}(r_{t,j})$, with $\bot$ denoting failure.
            \If{$x_{t,j}\neq\bot$ and $r_{t,j}\models C$}
                \State Score the product, $\mathbf{y}_{t,j}\gets f(x_{t,j})$.
                \State Store the valid candidate, $V_t\gets V_t\cup\{(r_{t,j},x_{t,j},\mathbf{y}_{t,j})\}$.
            \Else
                \State Store the failed route evidence, $F_t\gets F_t\cup\{r_{t,j}\}$.
            \EndIf
        \EndFor
    \EndIf
    \Statex \textcolor{green!50!black}{\(\triangleright\) Selection and memory}
    \State Select survivors, $P_{t+1,z_t}\gets\operatorname{Survive}(P_{t,z_t}\cup V_t)$.
    \State Update island memory, $M_{t+1,z_t}\gets\operatorname{UpdateMem}(M_{t,z_t},V_t,F_t)$.
    \State Update history, $H_{t+1}\gets\operatorname{UpdateHist}(H_t,V_t,F_t)$.
    \State Carry inactive islands forward, $P_{t+1,z},M_{t+1,z}\gets P_{t,z},M_{t,z}$ for $z\neq z_t$.
\EndFor
\State \Return $\operatorname{Best}(\{P_{N,z},M_{N,z}\}_{z\in\mathcal{Z}})$.
\end{algorithmic}
\end{algorithm}

\subsection{Variation operators}
\label{subsec:variation_operators_revised}

The strategy object $\sigma_t$ does not directly edit atoms or write a final
molecule. Instead, it gives soft preferences that the local sampler converts into
one of the route level variation operators illustrated in
Figs.~\ref{fig:create_new_route_control} -- \ref{fig:substitute_route_step_control}.
The LLM therefore selects the kind of route modification to emphasize, while
local code instantiates the concrete reaction templates, compatible reactants,
and executable reaction steps.

The action window $A_t$ provides the allowable chemistry for the current
iteration, while $\sigma_t$ provides weights over how that chemistry should be
used. These weights include route length, move type, route position, reaction
family, reaction slot, and building block preferences. The sampler interprets
them as a local policy over $A_t$: it first chooses which control to apply, then
fills that control with compatible templates and reactants from the action
window.

\paragraph{Operator selection.}
For each proposed candidate, the sampler uses the operator weights contained in
$\sigma_t$ to choose one of the enabled route edit operations. This choice
selects the route edit operation, not the final molecule. Once a branch is
chosen, all newly introduced steps are sampled locally from $\mathcal{S}$ by
selecting a reaction template from $\mathcal{R}$, supplying compatible reactants
from $\mathcal{B}$ or earlier intermediates, and executing the SMARTS transform.

We now describe each possible route level variation operator.
\begin{enumerate}
    \item \textbf{Create New Route.} This is the fully
    generative route move. No parent route is used; the sampler chooses a route
    length $L_t$ allowed by the task and builds
\begin{equation}
    r'=(s'_1,\ldots,s'_{L_t}), \qquad s'_i\in\mathcal{S}.
\end{equation}
    Every step is newly instantiated from available reaction templates and
    compatible building block SMILES, so this operator provides the strongest
    exploration of the current action window.

\begin{figure}[H]
\centering
\includegraphics[width=0.74\linewidth]{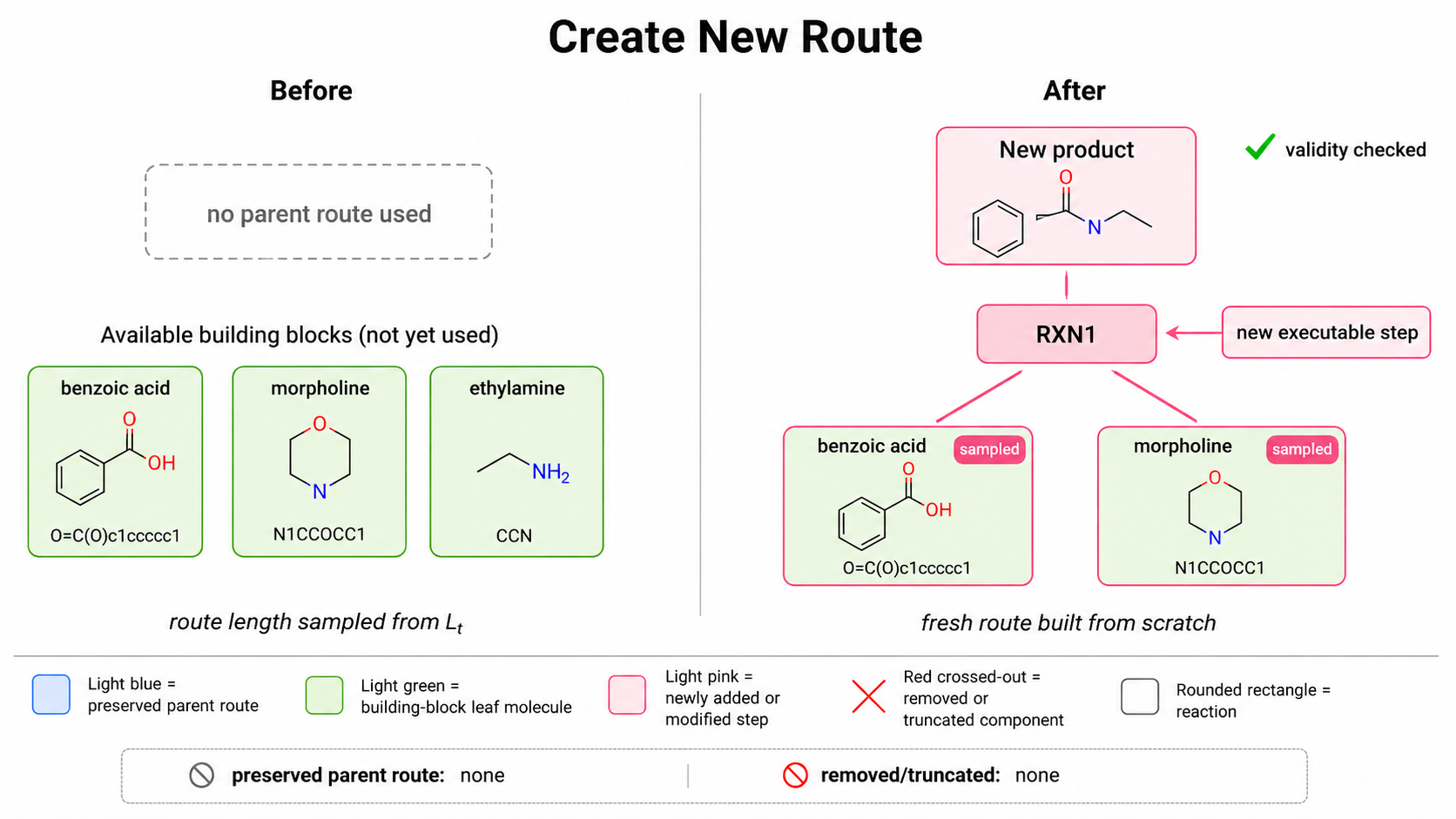}
\caption{\textbf{Create New Route.} This control ignores the parent
population and builds a fresh executable route from newly instantiated reaction
steps.}
\label{fig:create_new_route_control}
\end{figure}

    \item \textbf{Extend Parent.} This is an exploitative
    growth move. Given a parent route $r=(s_1,\ldots,s_L)$, the full parent is
    copied exactly and the sampler appends one or more new executable steps:
\begin{equation}
    (s_1,\ldots,s_L)
    \;\longrightarrow\;
    (s_1,\ldots,s_L,s'_{L+1},\ldots,s'_{L'}),
    \qquad s'_j\in\mathcal{S}.
\end{equation}
    This preserves the chemical pathway already found and searches for useful
    downstream chemistry, subject to the maximum route length.

\begin{figure}[H]
\centering
\includegraphics[width=0.74\linewidth]{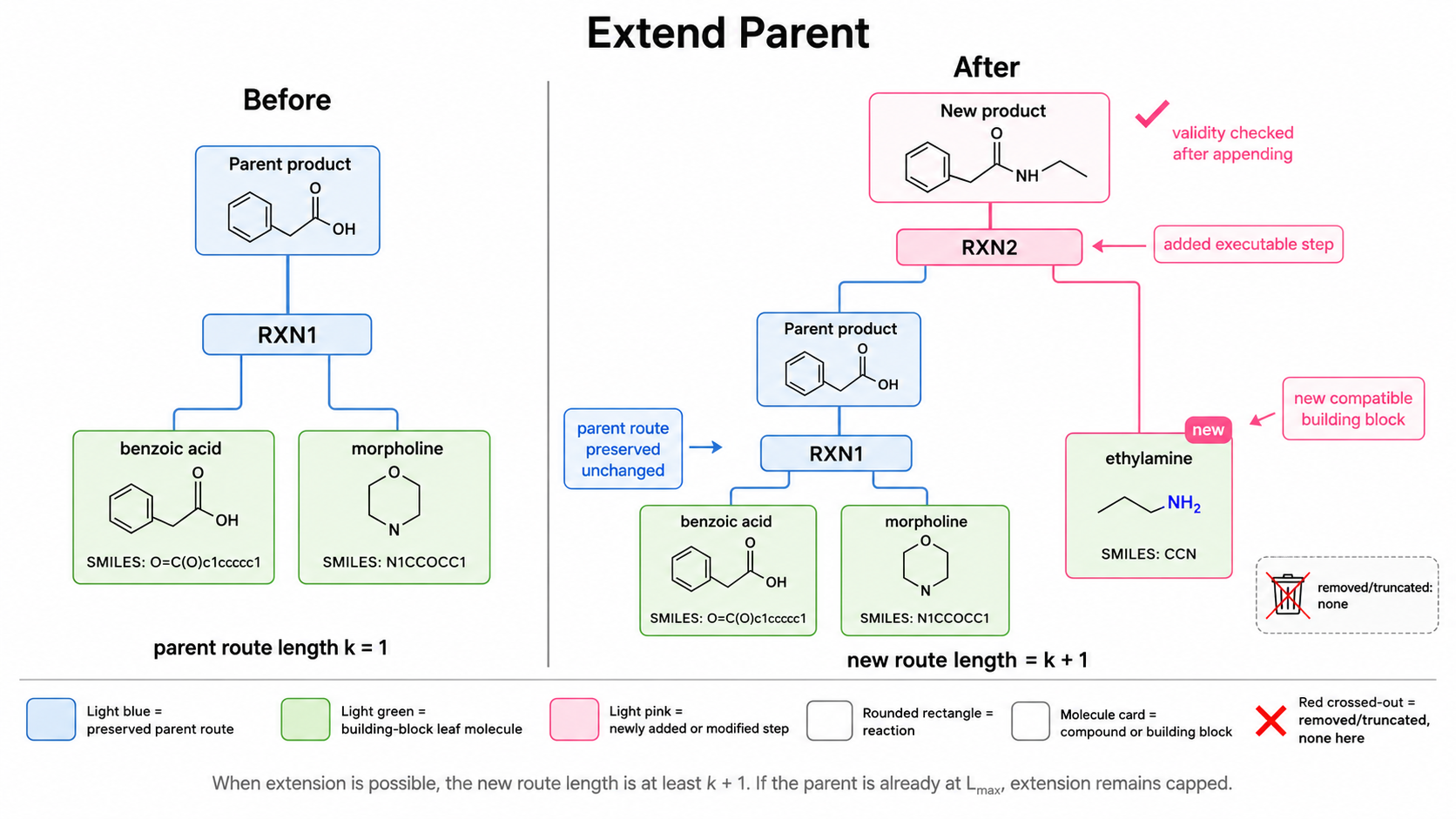}
\caption{\textbf{Extend Parent.} This control preserves the full parent
route and appends at least one new executable reaction step.}
\label{fig:extend_parent_control}
\end{figure}

    \item \textbf{Mutate Parent Prefix.} This is a local
    route repair or route variation move. A route position choice $h_t$
    determines the preserved prefix length $k$: an early mutation keeps little or
    none of the parent, a middle mutation keeps an intermediate prefix, and a
    final step mutation keeps most of the parent. The remaining suffix is
    discarded and regrown locally:
\begin{equation}
    (s_1,\ldots,s_L)
    \;\longrightarrow\;
    (s_1,\ldots,s_k,s'_{k+1},\ldots,s'_{L'}),
    \qquad s'_j\in\mathcal{S}.
\end{equation}
    The discarded suffix is not edited in place. Each replacement step must
    again pass SMARTS matching, reactant compatibility, RDKit execution, and
    route level validation.

\begin{figure}[H]
\centering
\includegraphics[width=0.74\linewidth]{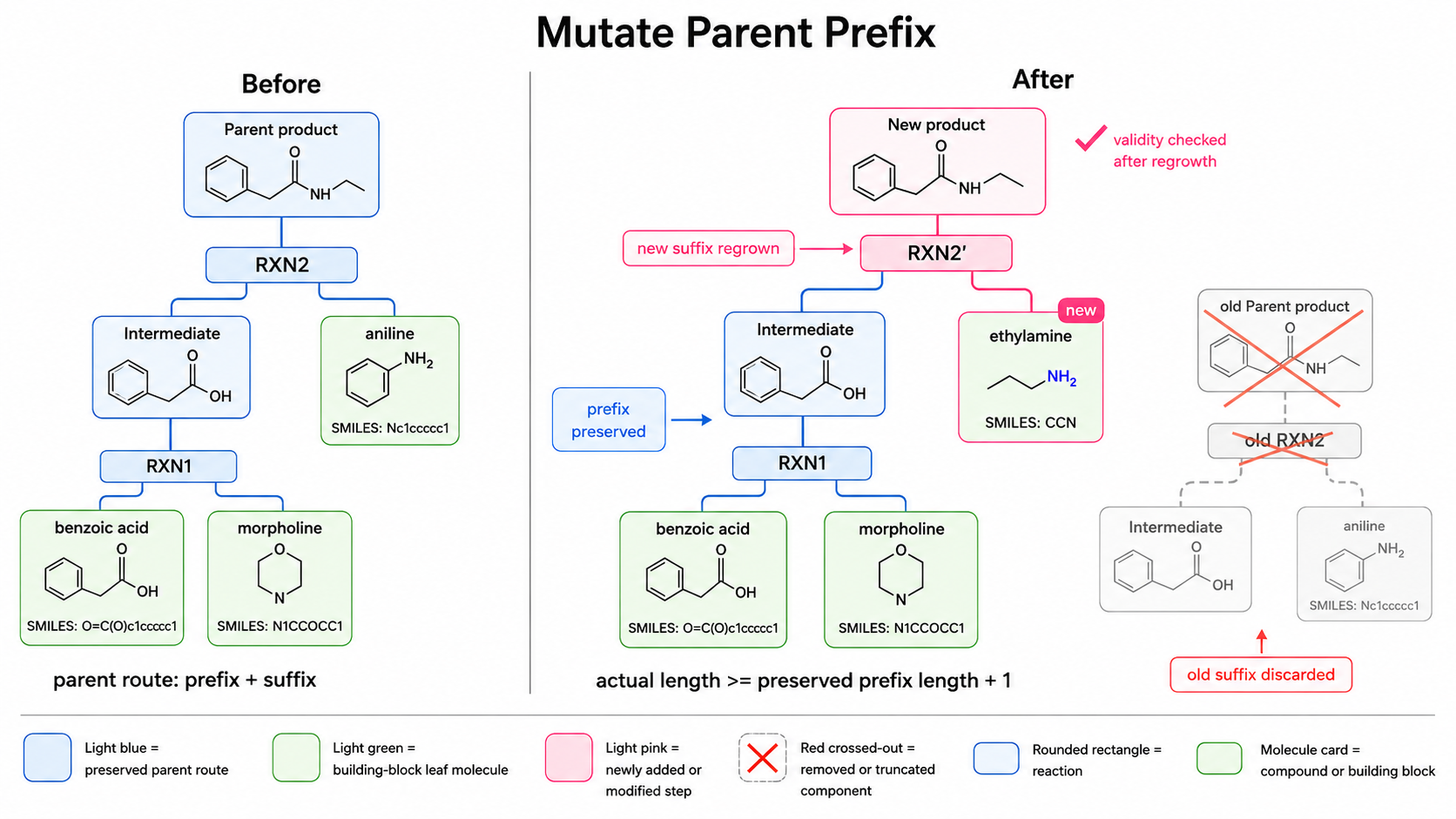}
\caption{\textbf{Mutate Parent Prefix.} This control preserves a
prefix of the parent route, discards the suffix, and regrows the remainder using
locally compatible SMARTS/SMILES choices.}
\label{fig:mutate_parent_prefix_control}
\end{figure}

    \item \textbf{Insert Route Step.} This is a route depth
    expansion move. Given a parent route, the sampler chooses an insertion
    position, keeps the unaffected part of the route, and inserts a new
    executable reaction layer. Because the inserted intermediate changes the
    product side connection, the adjacent step may also be resampled so the full
    route remains executable:
\begin{equation}
    (s_1,\ldots,s_i,s_{i+1},\ldots,s_L)
    \;\longrightarrow\;
    (s_1,\ldots,s_i,s'_{i+1},s'_{i+2},\ldots,s'_{L+1}),
    \qquad s'_j\in\mathcal{S}.
\end{equation}
    This operator is useful when the search needs an extra synthetic handle
    rather than a completely new route.

\begin{figure}[H]
\centering
\includegraphics[width=0.74\linewidth]{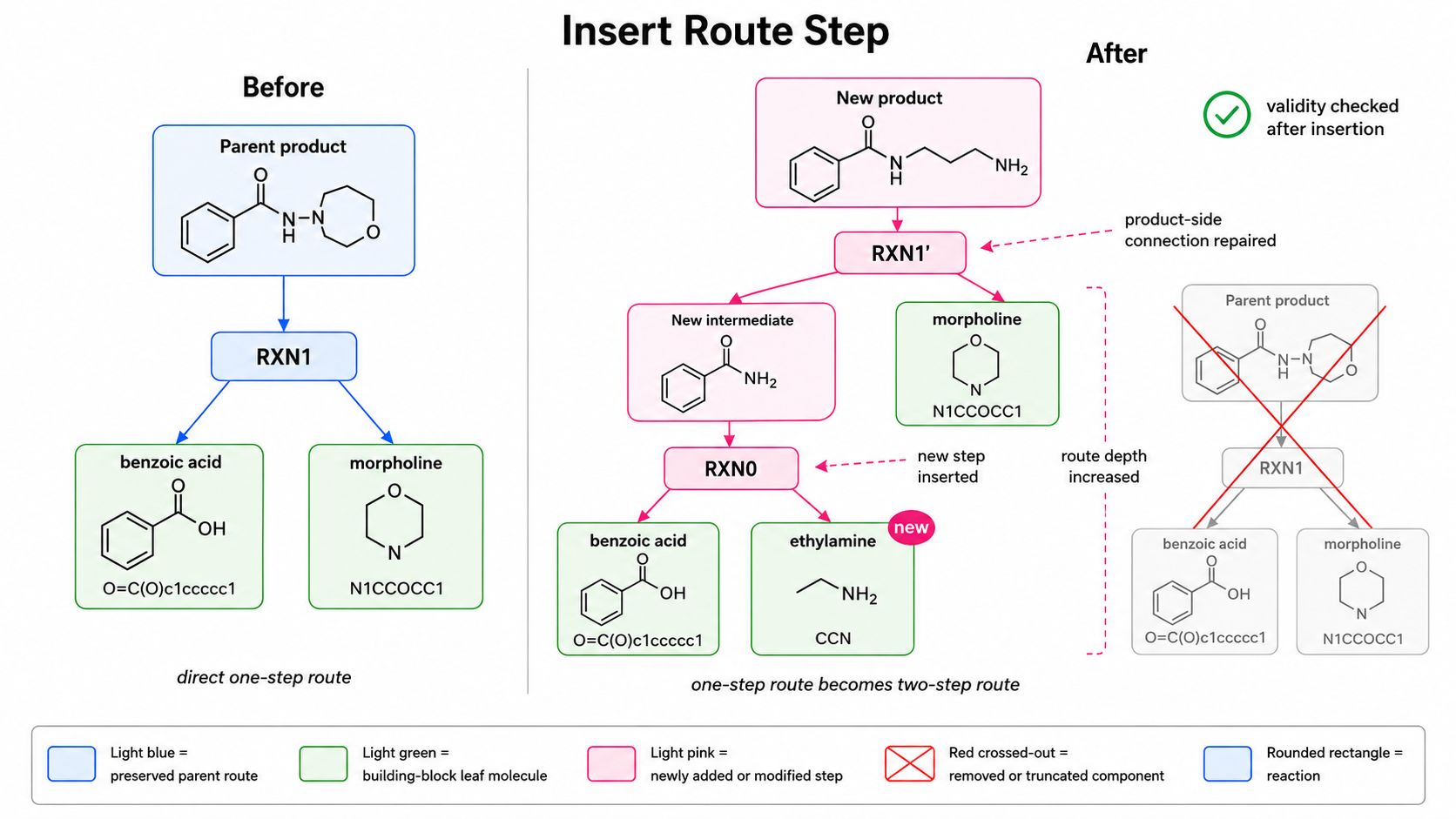}
\caption{\textbf{Insert Route Step.} This control inserts a new
reaction layer inside a parent route and repairs the affected product side
connection before validation.}
\label{fig:insert_route_step_control}
\end{figure}

    \item \textbf{Delete Route Step.} This is a route shortening
    move. The sampler removes one selected route step and keeps the remaining
    executable subtree:
\begin{equation}
    (s_1,\ldots,s_i,s_{i+1},\ldots,s_L)
    \;\longrightarrow\;
    (s_1,\ldots,s_i,\ldots,s_{L-1}).
\end{equation}
    No new reaction is added by this operator. It tests whether a simpler route
    can still give a valid product after the removed layer is discarded.

\begin{figure}[H]
\centering
\includegraphics[width=0.74\linewidth]{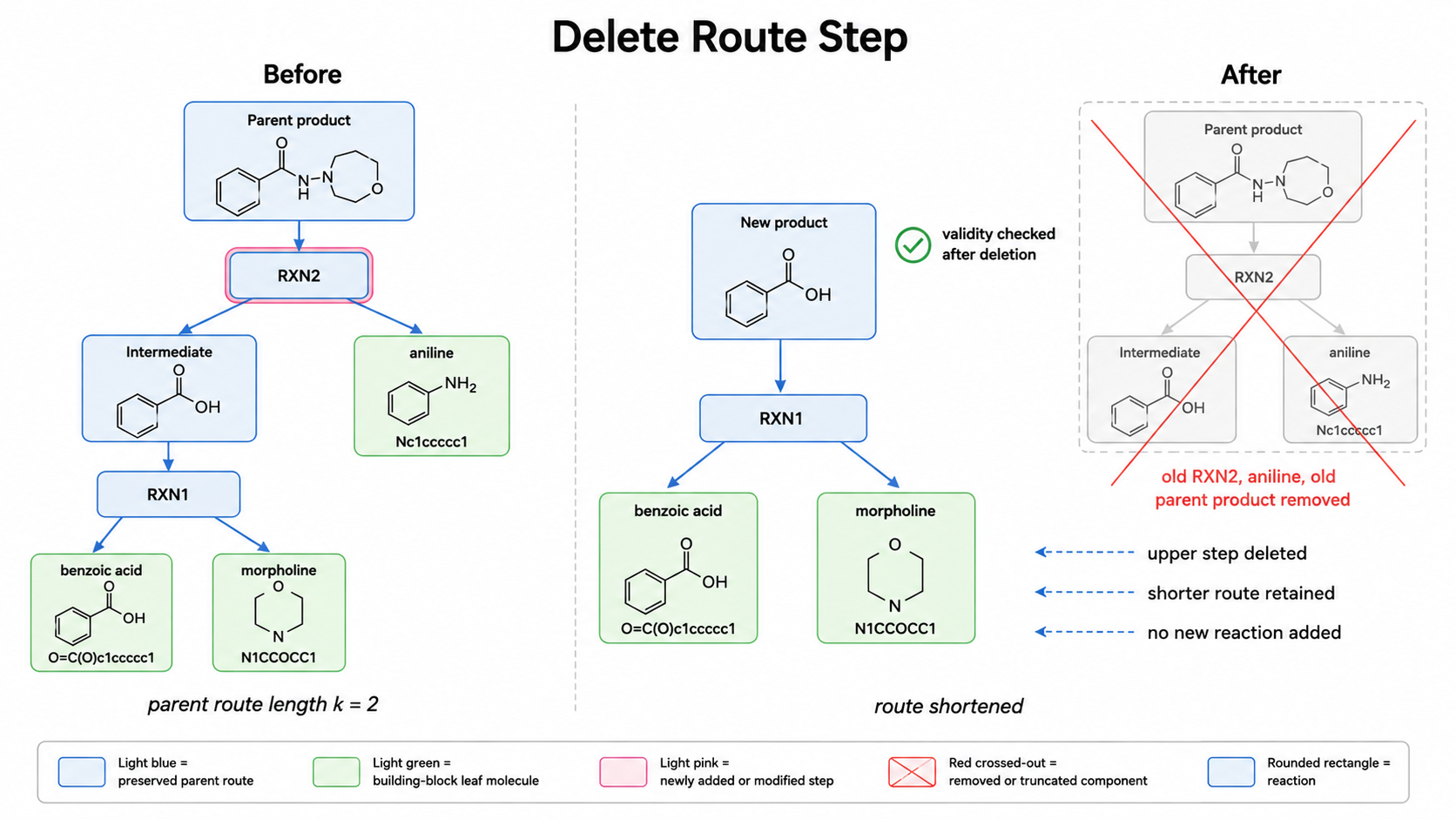}
\caption{\textbf{Delete Route Step.} This control removes a route
step, retains the remaining executable subtree, and validates the shorter
route.}
\label{fig:delete_route_step_control}
\end{figure}

    \item \textbf{Substitute Building Block.} This is the most local
    reactant level move. The reaction context is preserved, but one
    building block leaf is replaced by a new compatible building block sampled
    from $\mathcal{B}$:
\begin{equation}
    s_i=(R_i,b_1,\ldots,b_m)
    \;\longrightarrow\;
    s'_i=(R_i,b_1,\ldots,b'_q,\ldots,b_m),
    \qquad b'_q\in\mathcal{B}.
\end{equation}
    The reaction template stays fixed, but the product may change because one
    reactant changed. The substituted building block must still satisfy the
    required reaction slot.

\begin{figure}[H]
\centering
\includegraphics[width=0.74\linewidth]{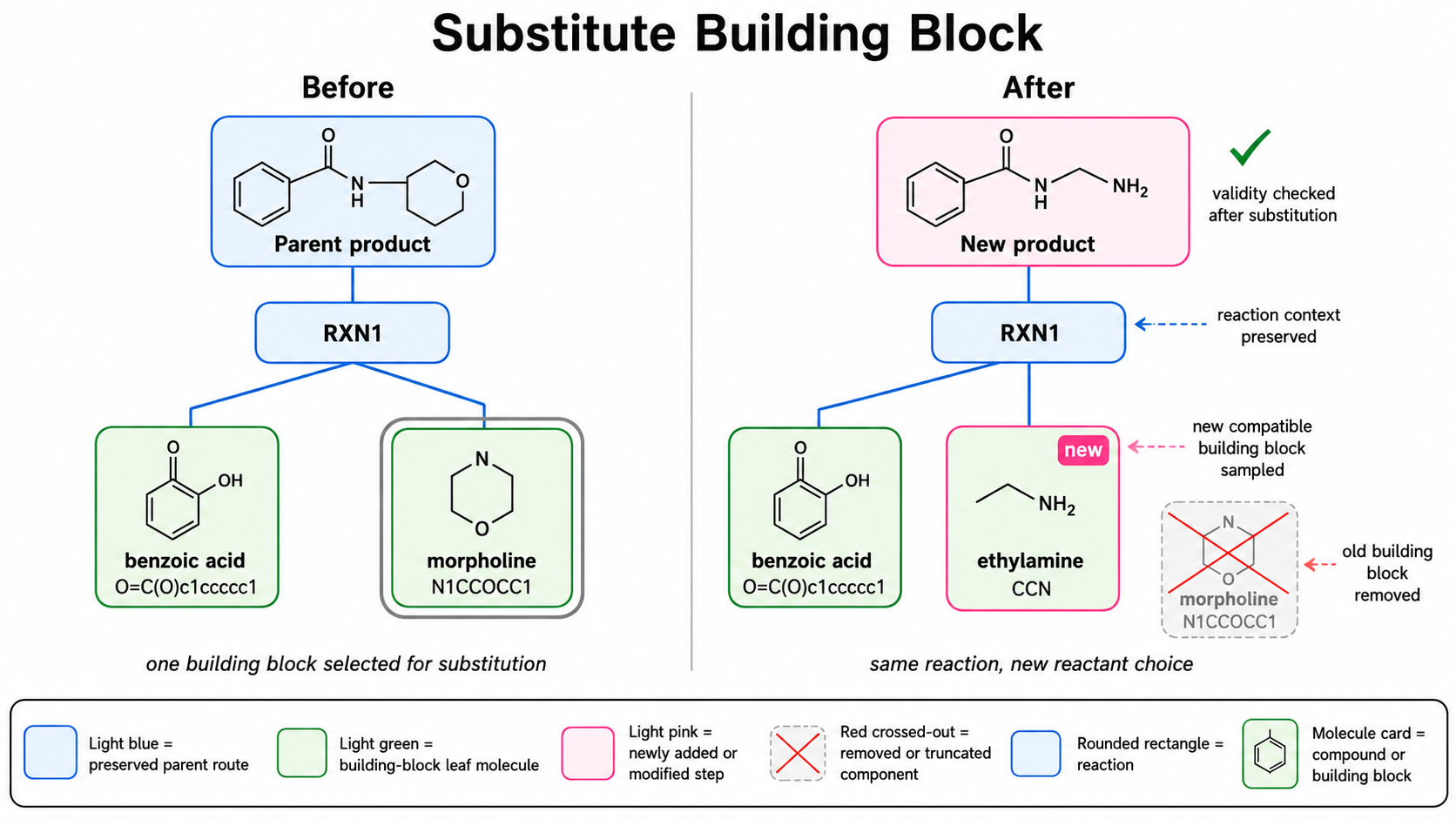}
\caption{\textbf{Substitute Building Block.} This control keeps the
reaction context and replaces one building block input with a newly sampled
compatible building block.}
\label{fig:substitute_building_block_control}
\end{figure}

    \item \textbf{Substitute Route Step.} This is a step level
    replacement move. The sampler selects one step, preserves the unaffected
    prefix, and replaces the selected reaction with a newly sampled compatible
    step. If the replaced step changes an intermediate needed downstream, the
    affected suffix is repaired or regrown:
\begin{equation}
    (s_1,\ldots,s_i,s_{i+1},\ldots,s_L)
    \;\longrightarrow\;
    (s_1,\ldots,s_{i-1},s'_i,s'_{i+1},\ldots,s'_{L'}),
    \qquad s'_j\in\mathcal{S}.
\end{equation}
    This operator changes a larger part of the route than a building block
    substitution, but still reuses any valid prefix chemistry from the parent.

\begin{figure}[H]
\centering
\includegraphics[width=0.74\linewidth]{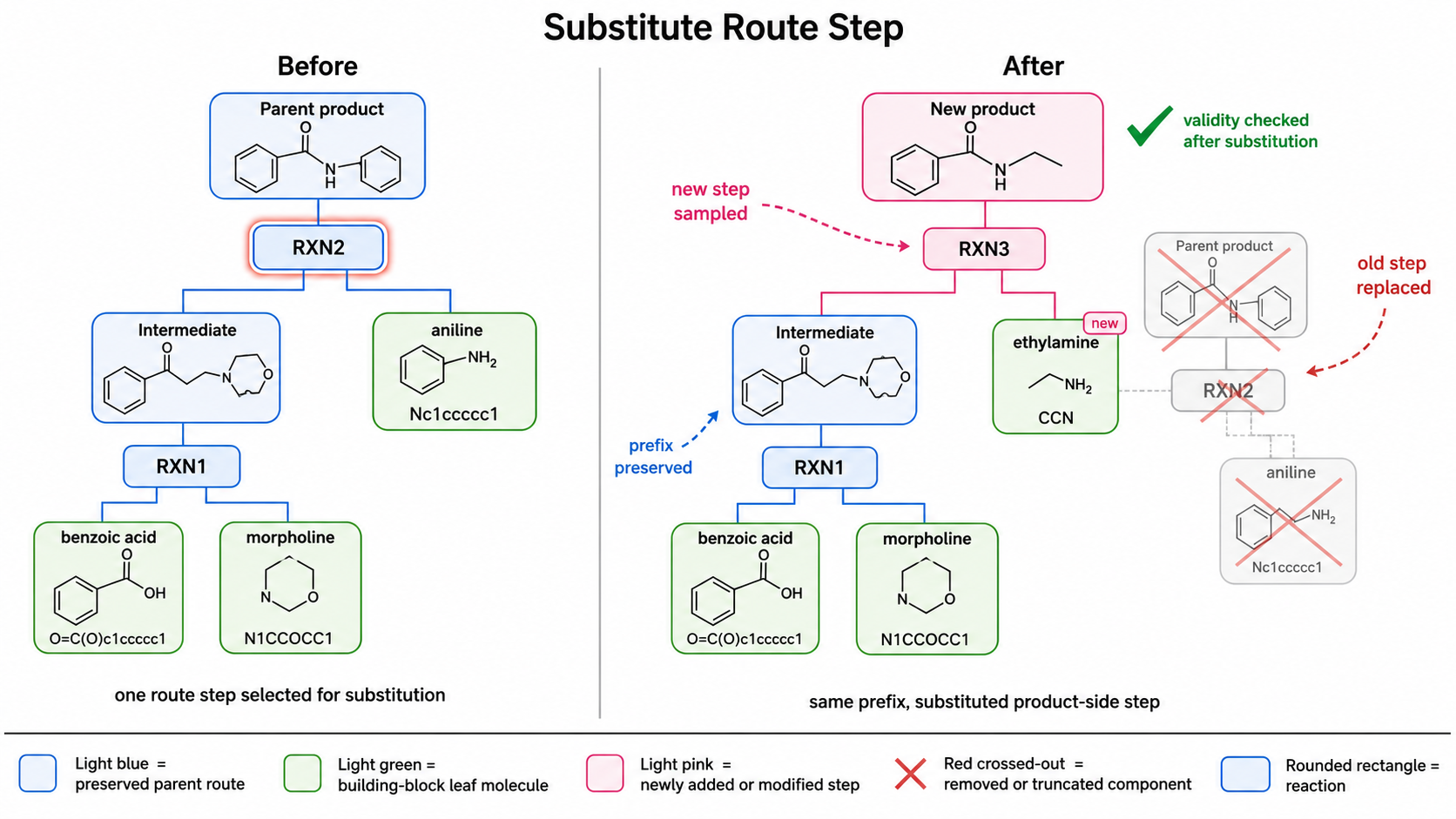}
\caption{\textbf{Substitute Route Step.} This control preserves the
unaffected prefix, replaces one route step, and repairs or regrows the affected
downstream connection.}
\label{fig:substitute_route_step_control}
\end{figure}
\end{enumerate}

\paragraph{Route length reconciliation.}
The prefix preserving control applies an additional route length reconciliation.
If $k$ steps are kept, the final route must contain at least one newly sampled
step after that prefix:
\begin{equation}
    L_{\mathrm{actual}}
    =
    \min\left(L_{\max},\,
    \max\left(L_t,\, k+1\right)\right).
\end{equation}
Thus, a sampled short route length cannot delete a preserved prefix, and a
parent based move remains bounded by the task maximum route length.

Additional enabled operators follow the same principle. They may replace a
terminal reaction, insert or delete a route step, mutate product side chemistry,
or substitute building blocks, but they still return only a proposed route. The
proposal is accepted only after the deterministic executor verifies every step
and the scorer evaluates the final product molecule.

\paragraph{Strategy arms as operator controllers.}
The controls above are applied through named
strategy arms. A strategy object may contain a set of arms
\(\mathcal{A}_t=\{a_1,\ldots,a_m\}\), where each arm has a budget fraction
\(\beta_a\) and a local control profile. The implementation normalizes these
fractions into candidate quotas and samples each arm separately:
\begin{equation}
    b_a
    =
    \left\lfloor
    b\,\frac{\max(\beta_a,0)}
    {\sum_{a'\in\mathcal{A}_t}\max(\beta_{a'},0)}
    \right\rceil .
\end{equation}
Each arm inherits the global strategy fields but may override controls in
\(\sigma_t\): route length weights, move type weights, route position weights,
preferred or avoided reaction templates, preferred reaction slots, and preferred
building blocks. These controls are interpreted only relative to the action
window \(A_t\); for example, a preferred reaction must be one of the templates
shown in \(A_t\). For a candidate assigned to arm \(a\), the sampler uses the
arm specific strategy controls, denoted here by \(\sigma_t^a\), and dispatches
to the corresponding variation operator described above. In this sense, arms
control how the fixed operator set is used; they do not create new chemistry
operators. The generated candidate stores the arm name, arm role, budget
fraction, and selected move type, which lets the reflection step report which
arms and operators produced valid, high scoring, duplicate, or failed routes.

\subsection{Fitness and memory}
\label{subsec:fitness_assessment_revised}

After route execution, each valid product molecule is evaluated by the configured
oracle. The scorer returns raw property values, normalized objective rewards,
and a scalar score
\begin{equation}
    F_{\mathrm{scalar}}(r)
    =
    \frac{\sum_i w_i\Phi_i(f_i(\mathbf{m}(r)))}{\sum_i w_i},
\end{equation}
where $\mathbf{m}(r)$ is the product obtained by executing route $r$, $f_i$ is a
property oracle, $\Phi_i$ maps the raw property to a bounded reward, and $w_i$ is
the objective weight. Multiobjective runs also retain the unaggregated objective
vector for Pareto selection. For the sEH task, the property vector contains the
sEH binding proxy, synthetic accessibility reward, and QED. Routes that fail
execution are marked invalid and stored as failures. Routes that execute but fail
hard molecular constraints or lack required objective values receive zero scalar
reward but can still be retained as informative examples.

To preserve population diversity and mitigate premature convergence, we split the 
population into islands, following the memory based evolutionary guidance strategy 
used in ~\cite{romera2024mathematical}, but with synthetic routes
as the stored objects. Each island $z$ maintains two related state variables. The
survivor population $P_{t,z}$ stores routes that can be selected as parents in
future generations. The memory $M_{t,z}$ stores a broader evidence buffer used
for prompting: top successes, randomly sampled successes, informative failures,
and recent low scoring valid routes. This separation lets survival remain
fitness driven while still exposing the LLM to negative and near miss evidence.

At iteration $t$, the active island is selected by a Boltzmann protocol over
island scores. Let $P^+_{t,z}\subseteq P_{t,z}$ denote the successful survivor
routes on island $z$, and define
\begin{equation}
    s_{t,z}
    =
    \begin{cases}
    |P^+_{t,z}|^{-1}\sum_{r\in P^+_{t,z}}F_{\mathrm{scalar}}(r),
    & |P^+_{t,z}|>0,\\
    0, & \text{otherwise}\footnotemark,
    \end{cases}
\end{equation}
\footnotetext{In early iterations, an island may contain evaluated routes but no
successful route yet. In that case, the score falls back to the mean scalar score
over all members of $P_{t,z}$; empty islands receive score zero.}
The probability of selecting island $z$ is
\begin{equation}
    \Pr(z_t=z)
    =
    \frac{\exp(s_{t,z}/\tau_z)}
    {\sum_{z'\in\mathcal{Z}}\exp(s_{t,z'}/\tau_{z'})},
\end{equation}
with temperature $\tau_z$ set by the island selection configuration. Larger
temperatures flatten the distribution and encourage exploration across islands;
smaller temperatures concentrate computation on islands whose recent successful
routes have higher mean fitness. With a single island, the same protocol reduces
to deterministic selection of that island.

Once an island is selected, parent routes are sampled from its successful
survivors by a score softmax rule. The prompt memory for the selected island is then assembled from the
island local buffer rather than from the full archive. Accepted offspring are registered back into the same
island, inactive islands carry their populations and memories forward, and the
global archive is used for logging, deduplication, and final analysis.

\subsection{Reflection and learning}
\label{subsec:memory_reflection_revised}

In reflective runs, each iteration has two local sampling phases. First, the LLM
emits a draft strategy and the local sampler evaluates a draft batch. The system
then summarizes the draft outcomes, including top candidates, failed routes,
low scoring routes, objective components, reaction outcomes, strategy arm
outcomes, and motif evidence. The LLM receives this summary and emits a revised
strategy for the remaining budget. The draft and revised batches are pooled for
archive logging, survivor selection, and memory update. A final learning report
summarizes which strategy arms, reaction families, motifs, and failure classes
should influence the next iteration.

Using the same proposal notation as above, the first pass samples a draft
strategy
\begin{equation}
    \sigma_t^0 \sim \pi_\theta(\cdot \mid p_t),
\end{equation}
which induces a draft route proposal distribution
\begin{equation}
    Q_t^0
    \sim
    K_{\sigma_t^0}(\cdot \mid A_t,P_{t,z_t},M_{t,z_t},H_t).
\end{equation}
Local execution and scoring convert the sampled draft batch
$B_t^0=\{r_{t,j}^0\}_{j=1}^{b_0}$ into an evidence summary
\begin{equation}
    E_t^0
    =
    S_{\mathrm{ref}}\!\left(
    B_t^0,\mathbf{m}(B_t^0),f(\mathbf{m}(B_t^0))
    \right).
\end{equation}
The reflection step conditions the next strategy on this observed evidence,
\begin{equation}
    \sigma_t^1
    \sim
    \pi_\theta(\cdot \mid p_t,\sigma_t^0,E_t^0),
\end{equation}
and the remaining budget is sampled from
\begin{equation}
    Q_t^1
    \sim
    K_{\sigma_t^1}(\cdot \mid A_t,P_{t,z_t},M_{t,z_t},H_t,E_t^0).
\end{equation}
The evaluated iteration batch is then
\begin{equation}
    B_t = B_t^0\cup B_t^1,
\end{equation}
which is the set passed to archive logging, survivor selection, memory update,
and learning report generation.

The reflection prompt explicitly tells the LLM that the observed evidence comes
from deterministic local execution and that the next response must be a revised
strategy, not a route or molecule. The operational prompt contains instructions
of the following form:
\begin{lstlisting}[style=mchprompt]
Within iteration first pass outcomes:
You already committed the initial strategy below in this same iteration.
Local code sampled, executed, scored, and deduplicated a draft batch.
Use this new evidence to correct or refine the strategy for the remaining
candidate budget.

Initial strategy JSON:
{ ... }

Observed first-pass summary JSON:
{
  "valid_count": 18,
  "invalid_count": 7,
  "best_score": 0.72,
  "top_scored_candidates": [...],
  "low_or_failed_candidates": [...],
  "reaction_outcomes": [...],
  "strategy_arm_outcomes": [...],
  "motif_scaffold_outcomes": [...]
}

Revision instructions:
1. Keep useful reaction families, slots, reactants, route lengths, or move types
   that produced higher scores.
2. Downweight or avoid patterns associated with invalid routes, duplicates,
   low scores, or poor objective tradeoffs.
3. Use failure_class, component scores, and strategy_arm outcomes to decide
   which arms deserve more or less budget.
4. Return one revised strategy JSON object. Do not output concrete routes,
   reactant lists, invented molecules, markdown, or prose.
\end{lstlisting}

After the pooled draft and revised batches have been scored and committed, the
system can ask for a compact learning report that will be included in the next
iteration's strategy prompt. This report is generated from the same deterministic
iteration summary, but its purpose is longitudinal memory rather than immediate
budget reallocation:
\begin{lstlisting}[style=mchprompt]
Role: You are writing a concise chemistry learning report for one
molecule-search iteration.
The report will be included in the next iteration's strategy prompt.
Write Markdown only. Be concrete, compact, and evidence-based.

Strategy JSON used this iteration:
{ ... }

Scored iteration summary:
{ ... }

Report requirements:
1. Summarize what chemical/reaction patterns seemed useful.
2. Summarize what looked harmful or low-scoring.
3. Summarize which motifs/scaffolds looked enriched, harmful, plateaued, or
   repairable using only observed deterministic annotations.
4. Recommend which persistent strategy arms to keep, retire, split, merge, or
   re-budget.
5. Suggest next-iteration strategy adjustments using exact reaction IDs or
   slots when supported.
6. State uncertainty; do not overclaim from weak evidence.
\end{lstlisting}

\begin{algorithm}[H]
\caption{Reflective within iteration strategy refinement}
\label{alg:reflection_refinement}
\begin{algorithmic}[1]
\Require $p_t,A_t,P_{t,z_t},M_{t,z_t},H_t,b_0,b_1,f,\pi_\theta$
\Ensure Pooled evaluated batch $B_t$ and learning evidence $E_t$
\State Sample draft strategy $\sigma_t^0\sim\pi_\theta(\cdot\mid p_t)$.
\State Sample draft routes $B_t^0\sim K_{\sigma_t^0}(\cdot\mid A_t,P_{t,z_t},M_{t,z_t},H_t)$.
\State Execute and score $B_t^0$ locally.
\State Build evidence $E_t^0\gets S_{\mathrm{ref}}(B_t^0,\mathbf{m}(B_t^0),f(\mathbf{m}(B_t^0)))$.
\State Sample revised strategy $\sigma_t^1\sim\pi_\theta(\cdot\mid p_t,\sigma_t^0,E_t^0)$.
\State Sample revised routes $B_t^1\sim K_{\sigma_t^1}(\cdot\mid A_t,P_{t,z_t},M_{t,z_t},H_t,E_t^0)$.
\State Execute and score $B_t^1$ locally.
\State Pool batches, $B_t\gets B_t^0\cup B_t^1$.
\State Build iteration evidence $E_t\gets S_{\mathrm{ref}}(B_t,\mathbf{m}(B_t),f(\mathbf{m}(B_t)))$.
\State Generate the learning report from $E_t$ for inclusion in future prompts.
\end{algorithmic}
\end{algorithm}

The resulting loop keeps the roles explicit. The LLM plans the search and
reflects on evidence; local chemistry tools generate, validate, execute, score,
deduplicate, and store routes. The objective values and losses used for
selection are therefore grounded in deterministic route execution and quantitative
oracles, not in LLM self assessment.

\addsec{Acknowledgements}
We thank Nikhil Abhyankar for helpful discussions and guidance related to his
work on evolutionary materials design workflows. We also
acknowledge the High Performance Computing department at the University of
Potsdam for computational support. D.A.F supported by the DOE under Award Number DOE-SC0010008.

\bibliography{acs-latex-template}

\appendix
\section{Objective and Scoring Details}
\label{app:objective_scoring}

This appendix gives the implementation level scoring details that are omitted
from the main Methods. The main text treats the objective as a blackbox fitness
function; here we specify the normalization and scalar aggregation used by the
implementation.

\subsection{Scalar objective rewards}

For a valid executed route $r$, the scorer computes raw objective values
$f_i(\mathbf{m}(r))$. Each objective is converted to a bounded reward $\Phi_i$
using its configured direction and bounds. For a maximize objective with bounds
$(l_i,u_i)$,
\begin{equation}
    \Phi_i(f_i(\mathbf{m}(r))) =
    \operatorname{clip}\left(\frac{f_i(\mathbf{m}(r))-l_i}{u_i-l_i},0,1\right).
\end{equation}
For a minimize objective,
\begin{equation}
    \Phi_i(f_i(\mathbf{m}(r))) =
    \operatorname{clip}\left(\frac{u_i-f_i(\mathbf{m}(r))}{u_i-l_i},0,1\right).
\end{equation}
For a target objective with target $\mu_i$ and scale $s_i$,
\begin{equation}
    \Phi_i(f_i(\mathbf{m}(r))) =
    \frac{1}{1+\lvert f_i(\mathbf{m}(r))-\mu_i\rvert/s_i}.
\end{equation}
The scalar score is the weighted mean
\begin{equation}
    F_{\mathrm{scalar}}(r)
    =
    \frac{\sum_i w_i\Phi_i(f_i(\mathbf{m}(r)))}
    {\sum_i w_i}.
\end{equation}

\subsection{sEH objective components}

The sEH configuration optimizes three objectives with equal weights:
\begin{equation}
    \big(
    \mathrm{sEH}(\mathbf{m}(r)),\;
    \mathrm{SA}_{\mathrm{norm}}(\mathbf{m}(r)),\;
    \mathrm{QED}(\mathbf{m}(r))
    \big).
\end{equation}
The sEH term is returned by the pretrained binding proxy and normalized to
$[0,1]$. The synthetic accessibility reward is derived from the RDKit SA score as
\begin{equation}
    \mathrm{SA}_{\mathrm{norm}}(\mathbf{m}(r))
    =
    \operatorname{clip}\left(\frac{10-\mathrm{SA}_{\mathrm{raw}}(\mathbf{m}(r))}{9},0,1\right),
\end{equation}
so larger values indicate easier synthesis. QED is the raw RDKit
drug likeness score in $[0,1]$ for the NSGA-II sEH configuration. The scalar
logging score for equal weights is
\begin{equation}
    F_{\mathrm{scalar}}(r)
    =
    \frac{
    \mathrm{sEH}(\mathbf{m}(r))+
    \mathrm{SA}_{\mathrm{norm}}(\mathbf{m}(r))+
    \mathrm{QED}(\mathbf{m}(r))
    }{3}.
\end{equation}

\subsection{Search losses}

The evolutionary loop does not train the LLM with gradients. For analysis, the
blackbox loss associated with a route can be written as
\begin{equation}
    \mathcal{L}_{\mathrm{search}}(r)=-F_{\mathrm{scalar}}(r),
\end{equation}
or componentwise as
\begin{equation}
    \ell_i(r)=1-\Phi_i(f_i(\mathbf{m}(r))).
\end{equation}
These quantities are derived from oracle evaluations and are used for ranking,
reflection summaries, and failure analysis.

\end{document}